\documentclass[aip,amsmath,amssymb,reprint]{revtex4-1}
\pdfoutput=1
\usepackage{amsmath,amsfonts,amssymb,bm,graphicx,hyperref}
\usepackage{color}

\newcommand{\be}{\begin{equation}}
\newcommand{\ee}{\end{equation}}
\newcommand{\SC}{S_{\rm conf}}
\definecolor{darkblue}{rgb}{0,0,0.6}
\DeclareMathAlphabet\mathrsfso      {U}{rsfso}{m}{n}

\newcommand{\res}[1]{{\color{black} #1}}
\hypersetup{bookmarksopen=true,
pdftitle="",
pdfauthor="",  
pdftoolbar=false,                % toolbar hidden
pdfstartview={FitH},             % fits the width of the page to the window
pdfmenubar=true,                 %menubar shown
pdfhighlight=/O,                 %effect of clicking on a link
colorlinks=true,                 %couleurs sur les liens hypertextes
urlcolor=darkblue,
citecolor=darkblue,           
linkcolor=darkblue}

\begin{document}

\title{Configurational entropy of glass-forming liquids}

\author{Ludovic Berthier}

\author{Misaki Ozawa}

\author{Camille Scalliet}

\affiliation{Laboratoire Charles Coulomb (L2C), Universit\'e de 
Montpellier, CNRS, Montpellier, France}

\begin{abstract}
The configurational entropy is one of the most important thermodynamic quantities characterizing supercooled liquids approaching the glass transition. Despite decades of experimental, theoretical, and computational investigation, a widely accepted definition of the configurational entropy is missing, its quantitative characterization remains fraud with difficulties, misconceptions and paradoxes, and its physical relevance is vividly debated. Motivated by recent computational progress, we offer a pedagogical perspective on the configurational entropy in glass-forming liquids. We first explain why the configurational entropy has become a key quantity to describe glassy materials, from early empirical observations to modern theoretical treatments. We explain why practical measurements necessarily require approximations that make its physical interpretation delicate. We then demonstrate that computer simulations have become an invaluable tool to obtain precise, non-ambiguous, and experimentally-relevant measurements of the configurational entropy. We describe a panel of available computational tools, offering for each method a critical discussion. This perspective should be useful to both experimentalists and theoreticians interested in glassy materials and complex systems.
\end{abstract}

\date{\today}

\maketitle

\section{Configurational entropy and glass formation}

\label{sec:introduction}

\subsection{The glass transition}

When a liquid is cooled, it can either form a crystal or avoid crystallization and become a supercooled liquid. In the latter case, the liquid remains structurally disordered, but its relaxation time increases so fast that there exists a temperature, called the glass temperature $T_{\rm g}$, below which structural relaxation takes such a long time that it becomes impossible to observe. The liquid is then trapped virtually forever in one of many possible structurally disordered states: this is the basic phenomenology of the glass transition.~\cite{RMP,EAN96,DS01,cavagna2009supercooled} Clearly, $T_{\rm g}$ depends on the measurement timescale and shifts to lower temperatures for longer observation times. The experimental glass transition is not a genuine phase transition, as it is not defined independently of the observer. 

The rich phenomenology characterizing the approach to the glass transition has given rise to a thick literature. It is not our goal to review it, and we refer instead to previous articles.~\cite{RMP,EAN96,DS01,cavagna2009supercooled,dyre2006colloquium,binder2011glassy,ediger2012perspective,langer2014theories,edigertoday} There are convincing indications that the dynamic slowing down of supercooled liquids is accompanied by an increasingly collective relaxation dynamics. This is seen directly by the measurement of growing lengthscales for these dynamic heterogeneities,~\cite{BBBCS11,ediger2000spatially,karmakar2014growing} or more indirectly by the growth of the apparent activation energy for structural relaxation, as seen in its non-Arrhenius temperature dependence. These observations suggest an interpretation of the experimental glass transition in terms of a generic, collective mechanism possibly controlled by a sharp phase transition.~\cite{Ta11} `Solving the glass problem' thus amounts to identifying and obtaining direct experimental signatures about the fundamental nature and the mathematical description of this mechanism.

\begin{figure}[b]
\includegraphics[width=0.49\columnwidth]{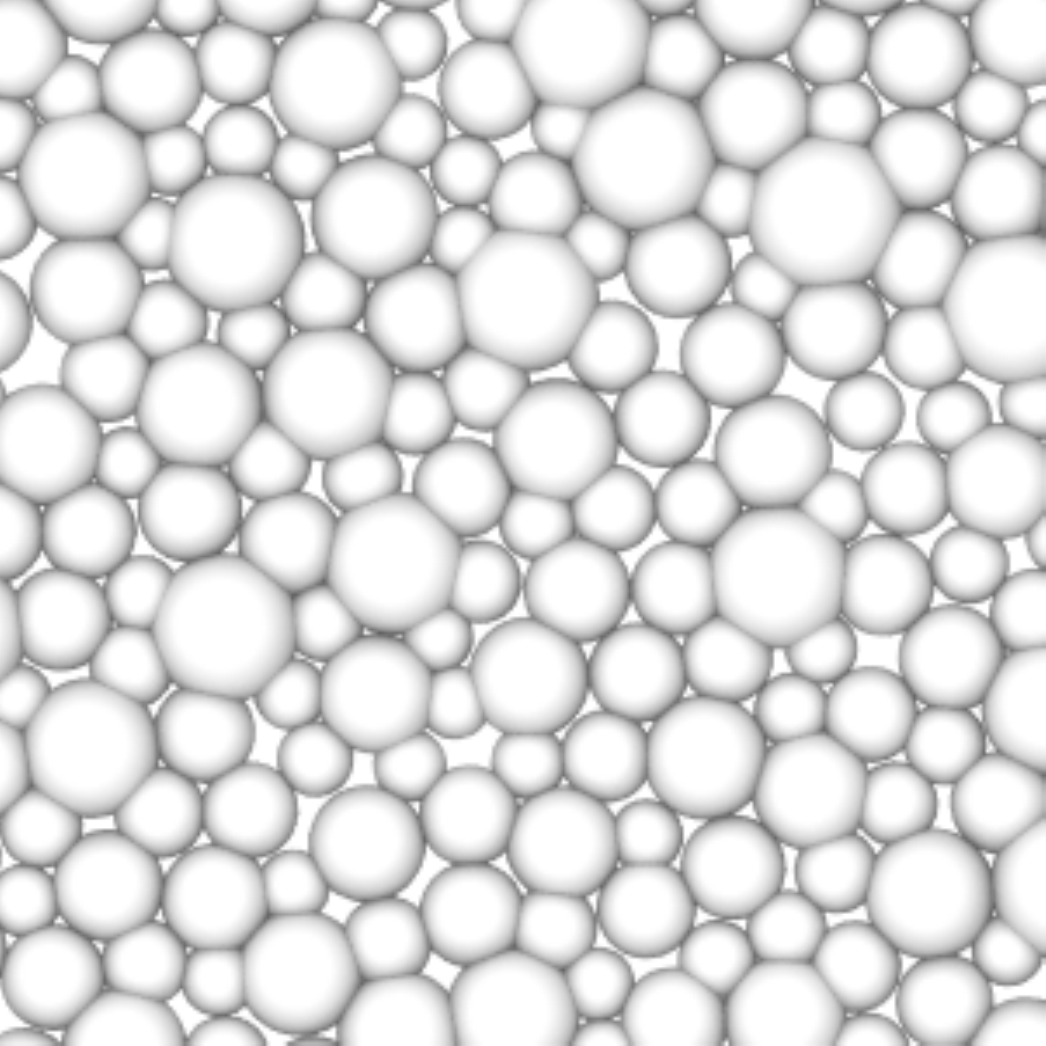}
\includegraphics[width=0.49\columnwidth]{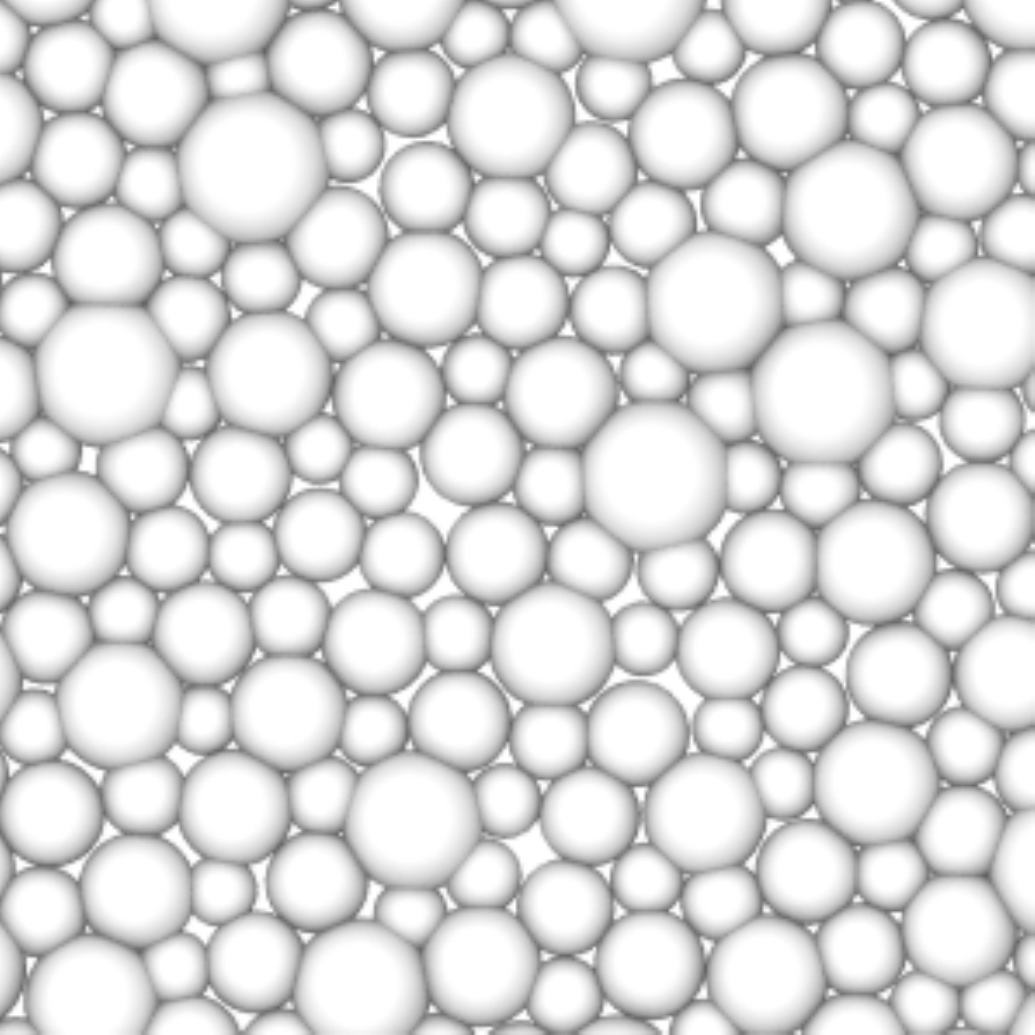}
\caption{Two equilibrium configurations of a two-dimensional glass-forming model characterized by relaxation times that differ by a factor $10^{12}$. The two density profiles \res{appear to the naked eye similarly featureless.} These two states in fact differ by the number of available equilibrium states and the configurational entropy quantifies this difference.}
\label{fig:cartoon}
\end{figure}

Why is this endeavor so difficult as compared to other phase transformations encountered in condensed matter?~\cite{stanley1971phase,chaikin1995principles} The core problem is illustrated in Fig.~\ref{fig:cartoon} by two particle configurations taken from a recent computer simulation.~\cite{berthier2018zero} 
The left panel shows an equilibrium configuration of a two-dimensional liquid with a relaxation time of order $10^{-10} \, {\rm s}$, using experimental units appropriate for a molecular system. The right panel shows another equilibrium configuration now produced close to $T_{\rm g}$ with an estimated relaxation timescale of order $100 \, {\rm s}$. The system on the right flows $10^{12}$ times slower than the one on the left, but to the naked eye both configurations look quite similar. In conventional phase transitions,~\cite{stanley1971phase,chaikin1995principles} a structural change takes place and some form of (crystalline, nematic, ferromagnetic, etc.) order appears. Glass formation is not accompanied by such an obvious structural change. Therefore, the key to unlock the glass problem is to first identify the correct physical observables to distinguish between the two \res{configurations} in Fig.~\ref{fig:cartoon}.

Several theories, scenarios and models have been developed in this context.~\cite{AG65,KTW89,LW07,BB09,tarjus2005frustration,dyre2006colloquium,angell2008glass,mauro2009viscosity,chandler2010dynamics,tanaka2012bond,stillinger2013glass,mirigian2014elastically} Some directly focus on the rich dynamical behavior approaching the glass transition,~\cite{chandler2010dynamics} while others advocate some underlying phase transitions of various kinds,~\cite{AG65,LW07,BB09} possibly involving some `hidden' or amorphous order.

In this perspective, we explore one such research line, in which configurational entropy associated with a growing amorphous order plays the central role.~\cite{LW07,BB09,kirkpatrick2015colloquium} We argue that recent developments in computational techniques offer exciting perspectives for future work, allowing the determination of complex observables that are not easily accessible in experiments, as well as the exploration of temperature regimes relevant to experiments. 

\subsection{Why the configurational entropy?}

The fate of equilibrium supercooled liquids followed below $T_{\rm g}$ with inaccessibly long observation times was discussed 70 years ago by Kauzmann in a seminal work.~\cite{Kauz48} Since the supercooled liquid is metastable with respect to the crystal, Kauzmann compiled data for the excess entropy, $S_{\rm exc} \equiv S_{\rm liq}-S_{\rm xtal}$, where $S_{\rm liq}(T)$ and $S_{\rm xtal}(T)$ are the liquid and crystal entropies, respectively. 
Kauzmann observed that $S_{\rm exc}(T)$ decreases sharply with decreasing the temperature of the equilibrium supercooled liquid.

An extrapolation of the temperature evolution of $S_{\rm exc}$ from equilibrium data to lower temperatures suggests that $S_{\rm exc}$ becomes negative at a finite temperature, which led Kauzmann to comment:~\cite{Kauz48} {\it `Certainly it is unthinkable that the entropy of the liquid can ever be very much less than that of the solid.'} To avoid this paradoxical situation, referred to as the Kauzmann paradox or entropy crisis, he mentioned the possibility of a thermodynamic glass transition occurring well below $T_{\rm g}$, at a temperature now called the Kauzmann temperature, $T_K$. 
Although Kauzmann suggested that crystallization eventually prevents the occurrence of an entropy crisis, Kauzmann's intuition remains very influential, for good reasons. 

Gibbs and DiMarzio were the first to give theoretical insights to the temperature evolution of $S_{\rm exc}$, by analogy with a lattice polymer model whose entropy is purely configurational.~\cite{gibbs1956nature,GDM58} Hence the conventional name, `configurational entropy' and notation $S_{\rm conf}$, widely used in the experimental literature.~\cite{richert1998dynamics}
We show below that there is no, and that there cannot be any, unique definition of $S_{\rm conf}$. We nevertheless use the same symbol for all discussed estimates. In particular, $S_{\rm conf} \approx S_{\rm exc}$. 

\begin{figure}
\includegraphics[width=1.\columnwidth]{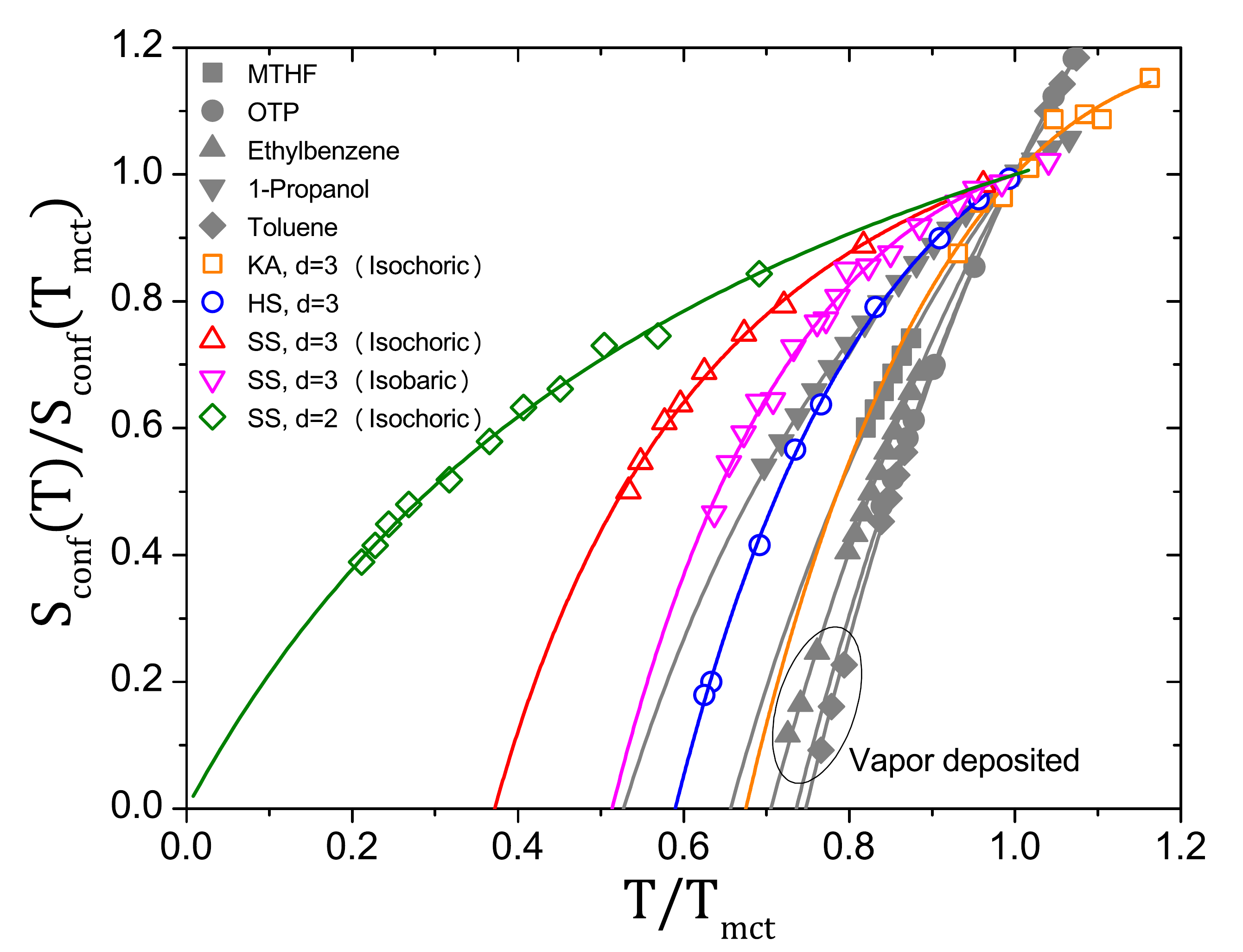}
\caption{Experimental and numerical determinations of the equilibrium configurational entropy in various models~\cite{ozawa2018configurational,berthier2018zero} and materials.~\cite{richert1998dynamics,tatsumi12,ediger2017perspective}
Data points extracted from vapor deposition experiments~\cite{ediger2017perspective} are indicated by the ellipse. Both axis are rescaled using the mode-coupling crossover as a reference temperature at which the relaxation time is about $10^{-7}$s. For hard spheres, the inverse of the reduced pressure, $1/p$, \res{replaces temperature.} Extrapolation to low temperatures suggests the possibility of an entropy crisis at a finite $T_K$ in $d=3$, whereas $T_K=0$ in $d=2$.}
\label{fig:kauzmann}
\end{figure}

We compile state-of-the-art experimental~\cite{richert1998dynamics,tatsumi12,ediger2017perspective} and numerical~\cite{ozawa2018configurational,berthier2018zero} data of $S_{\rm conf}$, and their extrapolation to low temperatures in Fig.~\ref{fig:kauzmann}. We employ a representation close to Kauzmann's original analysis,~\cite{Kauz48} rescaling $S_{\rm conf}$ by its value at some high temperature (we choose the mode-coupling temperature $T_{\rm mct}$,~\cite{gotze2008complex} for convenience).

In calorimetric experiments, the configurational entropy becomes constant below $T_{\rm g}$ upon entering the non-equilibrium glass regime, defining a residual entropy.~\cite{Kauz48,tatsumi12} The glass residual entropy is a non-equilibrium effect that has been extensively discussed.~\cite{kivelson1999metastable,goldstein2008reality,gupta2009configurational,johari2011entropy,schmelzer2018glass} Here, we focus on equilibrium supercooled liquids and do not discuss further the glass residual entropy and remove non-equilibrium measurements in Fig.~\ref{fig:kauzmann}.

The data for ethylbenzene and toluene are extended by combining conventional calorimetric measurements to data indirectly estimated from ultrastable glasses produced using vapor deposition.~\cite{Sw07,ediger2017perspective} \res{In that case, $T$ corresponds to the substrate temperature.} Various computational models using hard,~\cite{berthier2016equilibrium} soft,~\cite{NBC17} and Lennard-Jones potentials,~\cite{KA95a} along isochoric and isobaric paths, in spatial dimensions $d=2$~\cite{berthier2018zero} and $3$~\cite{ozawa2018configurational} are included along with experiments.~\cite{richert1998dynamics,tatsumi12,ediger2017perspective} 
%\MO{The relevant control parameter for the hard spheres is the reduced pressure $p=P/(\rho k_{\rm B}T)$, where $P$ and $\rho$ are the pressure and the number density. This natural control variables for the hard spheres plays a role akin to the inverse temperature in thermal liquids.~\cite{BW09}}
This representative data set demonstrates that all glass formers in dimension $d=3$ display a sharp decrease of $S_{\rm conf}$, even down to a temperature regime unavailable to Kauzmann. These results reinforce the idea that $S_{\rm conf}$ can vanish at a finite temperature, $T_K>0$. Simulation data in $d=2$ suggest instead that $S_{\rm conf}$ vanishes only at $T_K=0$, suggesting that a finite $T_K$ entropy crisis does not occur for $d<3$.~\cite{berthier2018zero} 

Of course, the data in Fig.~\ref{fig:kauzmann}
do not rule out the existence, at some yet inaccessible temperature, of a crossover in the behavior of $S_{\rm conf}$ that makes it smoothly vanish at $T=0$,~\cite{stillinger1988supercooled,debenedetti2003model} or remain finite with an equilibrium residual entropy in classical systems,~\cite{wolfgardt1996entropy,moreno2006non,donev2007configurational,smallenburg2013liquids,xu2016generalized} or a discontinuous jump due to an unavoidable crystallization,~\cite{Kauz48,tanaka2003possible,zanotto2017glassy} or a liquid-liquid transition,~\cite{angell2008glass} or a conventional (kinetic) glass transition.~\cite{schmelzer2018kauzmann} These alternative possibilities are not supported by data any better than the entropy crisis they try to avoid. It is impossible to comment on the many articles supporting the absence of a Kauzmann transition~\cite{stillinger1988supercooled,stillinger2001kauzmann,RT96,donev2007configurational,ZSMcK13,HSMcK03}, but we clarify below that none of them resists careful examination. The existence of a thermodynamic glass transition remains an experimentally and theoretically valid, but unproven, hypothesis.
Thus, extending configurational entropy measurements to even lower temperatures remains an important research goal.~\cite{royall2018race} 

As emphasized repeatedly, a negative $S_{\rm exc}$ is not prohibited by thermodynamic laws.~\cite{stillinger2001kauzmann} This is also not `unthinkable' since entropy is not a general measure of disorder. As a first counterexample, think of hard spheres for which the crystal entropy is larger than that of the fluid above the melting density under constant volume condition. A second example under constant pressure condition would be materials showing inverse melting.~\cite{doi:10.1063/1.1593018} 
A stronger reason to `resolve' the Kauzmann paradox is that if $S_{\rm liq}$ continues to decrease further below $S_{\rm xtal}$, the third law of thermodynamics could be violated.~\cite{kivelson1998kauzmann} However, the third law is conventionally interpreted as a consequence of the quantum nature of the system.~\cite{landau1980statistical} This implies that the Kauzmann paradox is not really problematic if considered within the realm of classical physics. In summary, {\it there is no theoretical need} to avoid the entropy crisis.

However, theoretical treatments rooted in Gibbs and DiMarzio's theory~\cite{gibbs1956nature,GDM58} relate the configurational entropy to the (logarithm of the) number of distinct glass states available to the system at a given temperature. A proper enumeration of those states must therefore result in a non-negative configurational entropy. In this interpretation, Fig.~\ref{fig:kauzmann} suggests that a fundamental change in the properties of the free-energy landscape must underlie glass formation. 

\begin{figure}
\includegraphics[width=0.99\columnwidth]{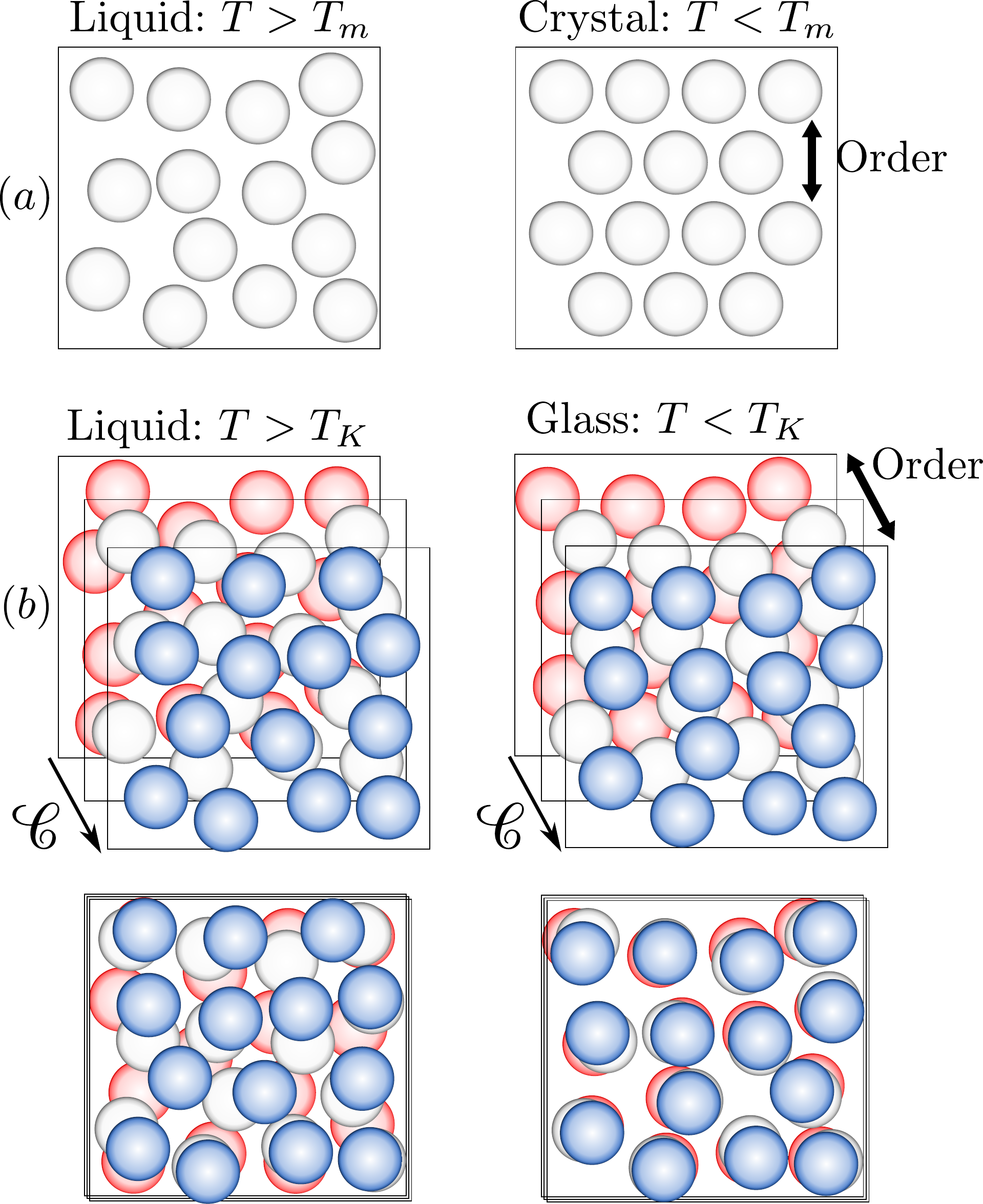}
\caption{(a) Crystallization at the melting temperature $T_{\rm m}$ corresponds to the emergence of periodic order in the density profile of a single configuration.
(b) The glass transition at $T_{\rm K}$ is detected by enumerating equilibrium configurations in configuration space $\mathrsfso{C}$. Glass order is revealed by comparing the degree of similarity (in practice, the overlap in Eq.~(\ref{eq:overlap})) of amorphous density profiles.}    
\label{fig:schematic_overlap}
\end{figure}  

A strong decrease of the configurational entropy answers the question raised by the \res{apparent structural similarity suggested} by the snapshots in Fig~\ref{fig:cartoon}. Conventional phase transitions deal with the `structure' of a single configuration,~\cite{stanley1971phase,chaikin1995principles} for instance the periodic order of the density profile for crystallization, see Fig.~\ref{fig:schematic_overlap}(a). By contrast, it is not the {\it nature} of the density profile that changes across the glass transition, but rather the {\it number} of distinct available profiles.~\cite{RMP} There are many distinct states available to the liquid, leading to a finite configurational entropy, but only a subextensive number in the putative thermodynamic glass phase, where $S_{\rm conf} = 0$. `Glass order' can thus only be revealed by the enumeration of equilibrium accessible states, see Fig.~\ref{fig:schematic_overlap}(b). 
%mention patch repetition length here?
 
A final general question is: how can a purely thermodynamic quantity be useful to understand slow dynamics? After all, the above phenomenological description of the glass transition relies on dynamics, and a connection to configurational entropy is not obvious. The first quantitative connection arose in 1965, when Adam and Gibbs proposed that the timescale for structural relaxation increases exponentially with $1/(TS_{\rm conf})$.~\cite{AG65} \res{Quantitatively, the modest decrease of $S_{\rm conf}(T)$ in Fig.~\ref{fig:kauzmann} could then be sufficient to account for the modest increase in the apparent activation energy, and for the large increase of relaxation times although this view remains heavily debated, to this day.~\cite{PhysRevLett.119.195501,doi:10.1063/1.5086509}}

Testing the Adam-Gibbs relation has become a straw man for a deeper issue:~\cite{dyre2009brief,richert1998dynamics,ZSMcK13,Mc2018} how can one (dis)prove the existence of a causal link between the rarefaction of equilibrium states and slow dynamics? In essence, the physical idea to be tested is that the driving force behind structural relaxation for $T > T_K$ is the configurational entropy gained by the system exploring distinct disordered states. Slower dynamics then arises when fewer states are available at lower $T$, since the system hardly finds new places to go. In this view, the two configurations in Fig.~\ref{fig:cartoon} relax at a much different rate not \res{because there structure is different,} but because much fewer equilibrium configurations are accessible to the configuration on the right. \res{This is indeed hard to recognize by the naked eye.}

\subsection{Mean-field theory of the glass transition}

\label{sec:MF}

Despite the diversity of theoretical work related to glass formation, the configurational entropy plays a central role. This is natural for theories rooted in thermodynamics and describe an entropy crisis,~\cite{AG65,KTW89,BB04} but theories based on a different mechanism must also explain the observed behavior of $S_{\rm conf}$, and the role played by a (possibly avoided) entropy crisis.~\cite{stillinger1988supercooled,tarjus2005frustration,angell2008glass} Finally, theories based on dynamics must explain why a rapidly changing $S_{\rm conf}$ is an irrelevant factor.~\cite{chandler2010dynamics,KGC13,BBT,CG} This makes the concept of configurational entropy, a careful understanding of its physical content, and the development of precise numerical measurements important research goals.

The first theory `predicting' an entropy crisis appeared about a decade after Kauzmann's work.~\cite{gibbs1956nature,GDM58} Inspired by lattice polymer studies,~\cite{flory1956statistical} Gibbs and DiMarzio identified the decrease of $S_{\rm exc}$ presented by Kauzmann with the reduction of the entropy computed within a set of mean-field approximations. In their lattice model, `states' were identified with microscopic configurations, with no need to subtract any vibrational contribution, $S_{\rm conf} \approx S_{\rm tot}$. An approximate statistical mechanics treatment of their model yields $S_{\rm tot} \to 0$ at a finite temperature. 

Revisions and extensions of the Gibbs-DiMarzio work abound.~\cite{gujrati1982lower,wittmann1991validity,wolfgardt1996entropy} Modern studies offer more detailed treatments of the polymer chain and refined approximations.~\cite{dudowicz2008generalized} The entropy may or may not vanish, depending on the approximations used and the ingredients entering the model.~\cite{xu2016generalized,xu2016entropy} 
An entropy crisis is thus not always present within the Gibbs-DiMarzio line of thought, but one cannot draw general conclusions about the existence of an entropy crisis in supercooled liquids. Moreover, the distinction between individual configurations and free-energy minima is generally not considered, which may be problematic.~\cite{BM00} Finally, these works rely heavily on the polymeric nature of the molecules to make predictions whose validity for simpler particle models or molecular systems is not guaranteed. 
These works nevertheless suggest that the presence of a Kauzmann transition could well be system-dependent. This is demonstrated by some specific colloidal models in which the entropy crisis is indeed avoided with a finite configurational entropy at zero temperature.~\cite{moreno2006non,smallenburg2013liquids} 

A coherent mean-field theory of the glass transition was recently derived for classical, off-lattice, point particle systems interacting with generic isotropic pair interactions.~\cite{mezard1999thermodynamics,PZ10,KPZ12,KPUZ13,CKPUZ16} The `mean-field' nature of the theory stems from the fact that it becomes mathematically exact in the limit of $d \to \infty$, whereas it amounts to an approximate analytic treatment for physical dimensions $d<\infty$. The nature of the glass transition found in this mean-field limit agrees with results obtained in the past in a variety of approximate treatments, starting with density functional theory of hard spheres,~\cite{singh1985hard} replica calculations of fully-connected spin glass models,~\cite{KT87,KT87b,KW87,KT88,KTW89,castellani2005spin} and others.~\cite{Derr81,BM01,RBMM04} 

The fact that distinct models and treatments yield similar results reflects a universal evolution of the free-energy landscape in glassy systems, with results rediscovered in a variety of contexts.~\cite{mezard2009information,kirkpatrick2015colloquium} The theory reveals the existence of sharp temperature scales where the topography of the free-energy landscape changes qualitatively. There exists a first temperature scale, $T_{\rm onset}$, above which a single global free energy minimum exists, and below which a large number, ${\cal N}$, of free-energy minima appear. This number scales exponentially with the system size, which allows for the definition of an entropy, $\Sigma  = \ln {\cal N}$,~\footnote{In this paper, we set the Boltzmann constant to unity.} also called complexity.
At a second temperature scale, $T_{\rm mct}  < T_{\rm onset}$, the partition function becomes dominated by those multiple free-energy minima. This transition shares many features with the dynamic transition first discovered in the context of mode-coupling theory.~\cite{gotze2008complex} The third critical temperature is $T_K < T_{\rm mct}$, below which the number of free-energy minima becomes subextensive, resulting in a vanishing complexity, $\Sigma(T \to T_K) \to 0$. 

An entropy crisis is thus an analytic result in mean-field theory, which provides a clear physical interpretation of the configurational entropy as the logarithm of the number of free-energy minima, $S_{\rm conf} \approx \Sigma = \ln {\cal N}$. A Kauzmann transition is exactly realized, and is referred to as a random first order transition (RFOT).  

The idea that the existence, number, and organization of distinct free-energy minima control the glass transition was elegantly captured by an approach developed by Franz and Parisi.~\cite{FP95,FP97} As in Landau theory%\MO{(It seems Ginsburg- means adding a gradient term.)}
, they expressed the free-energy, or effective potential $V(Q)$, as a function of a global order parameter $Q$. As illustrated in Fig.~\ref{fig:schematic_overlap}(b), the distinction between liquid and glass phases stems from the degree of similarity of particle configurations drawn from the Boltzmann distribution. Let us define an overlap, $Q$, as the degree of similarity of the density profiles of two equilibrium configurations, such that $Q \approx 0$ for uncorrelated profiles (liquid phase), and $Q \approx 1$ for similar profiles (glass phase); see Eq.~(\ref{eq:overlap}) below.

\begin{figure}
\includegraphics[width=0.95\columnwidth]{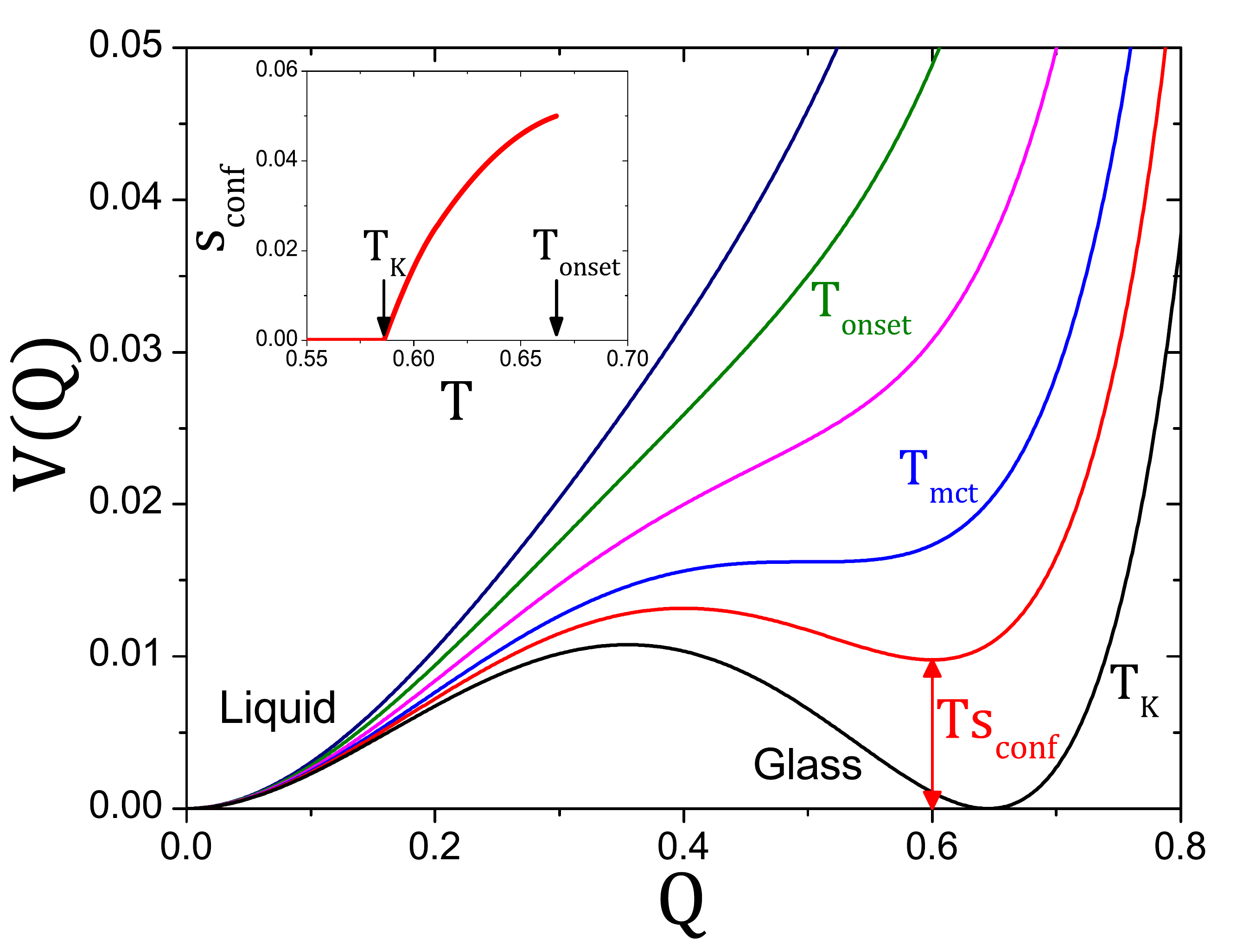}
\caption{Schematic plot of the Franz-Parisi free energy in mean-field theory. Inset: Temperature evolution of  the configurational entropy.}
\label{fig:FPMF}
\end{figure}  

The free-energy $V(Q)$ can be computed analytically for mean-field glass models, as shown in Fig.~\ref{fig:FPMF}. As expected, the global minimum of $V(Q)$ is near $Q \approx 0$ for $T>T_K$ as there \res{exist} so many distinct available states that two equilibrium configurations chosen at random have no similarity.
All critical temperatures mentioned above have a simple interpretation in this representation. The free-energy $V(Q)$ has non-convexity when $T<T_{\rm onset}$, it develops a secondary minimum when $T < T_{\rm mct}$, and this local minimum becomes the global one when $T$ reaches $T_K$. \res{The secondary minimum occurs for $Q$ slightly smaller than 1, due to thermal fluctuations.~\cite{BFP97}} In this description, mean-field glass theory shares similarities with ordinary first-order transitions.     

In the interesting regime, $T_{\rm K} < T < T_{\rm mct}$, the glass phase at high $Q$ is metastable with respect to the liquid phase at low $Q$. The free-energy difference between the liquid and glass phases results from confining the system within a restricted part of the configuration space. Preventing the system to explore the multiplicity of available free-energy minima entails an entropic loss, precisely given by the complexity, $T\Sigma (T)$. The temperature evolution of the configurational entropy $S_{\rm conf}$ is thus readily visualised and quantified from the Franz-Parisi free-energy as shown in Fig.~\ref{fig:FPMF}. The inset of Fig.~\ref{fig:FPMF} shows that a finite configurational entropy emerges discontinuously at $T_{\rm onset}$, and vanishes continuously at $T_K$.

The entropy crisis captured by the random first-order transition universality class is now validated by exact calculations performed in the large dimensional limit, $d \to \infty$.~\cite{CKPUZ16} This confers to RFOT a status similar to van der Waals theory for the liquid-gas transition. With its well-defined microscopic starting point, mean-field theory confirms that the configurational entropy is central to the understanding of supercooled liquids, and the rigorous treatment it offers puts phenomenological and approximate ideas introduced earlier by Kauzmann, Gibbs, DiMarzio, Adam, and others on a solid basis. This now serves as a stepping stone to describe finite dimensional effects.~\cite{effort1,effort2,effort3,effort4,effort5,effort6}

\subsection{Conceptual and technical problems}

\label{sec:conceptual}

Physically, the configurational entropy quantifies the existence of many distinct {\it glass states} that the system can access in equilibrium conditions. There are two main routes to measure $S_{\rm conf}$. 

First, one can subtract from the total entropy a contribution that comes from small thermal vibrations performed in the neighborhood of a given reference configuration: $S_{\rm conf} (T) \approx S_{\rm tot}(T) - S_{\rm glass} (T)$. In this view, $S_{\rm glass}(T)$ should be the entropy of an equilibrium system that does not explore distinct states at temperature $T$. This quantity can be measured straightforwardly in equilibrium for $T<T_K$, whereas some approximations are by construction needed to measure $S_{\rm glass}$ for $T>T_K$.

Experimentally, it is often assumed that $S_{\rm glass} \approx S_{\rm xtal}$, because it is possible to measure $S_{\rm xtal}$ in equilibrium using reversible thermal histories.~\cite{richert1998dynamics} This represents a well-defined and physically plausible proxy. It has been tested for some systems,~\cite{gold76,yamamuro1998calorimetric,J00,martinez2001thermodynamic,angell2002specific,smith2017separating} and its validity seems to be non-universal.~\cite{smith2017separating} 
%The outcome of this approach should therefore not be used to draw conclusions about the (ir)relevance of $S_{\rm conf}$ in supercooled liquids. 
We shall introduce in Sec.~\ref{sec:FL} a computational method to determine $S_{\rm glass}$ that makes no reference to the crystal.~\cite{frenkel1984new,coluzzi1999thermodynamics,AF07} 

%We note that even if the determination of $S_{\rm glass}$ in the liquid phase is not available, $S_{\rm tot}$ contains in principle all the information about the Kauzmann transition, as it would show a kink (associated to a jump in the specific heat) in equilibrium, as it transforms from $S_{\rm liq}$ to $S_{\rm glass}$ at $T_{\rm K}$. However, crossing $T_{\rm K}$ is generally impossible. Besides $S_{\rm tot}$ alone does not tell the location of the putative $T_{\rm K}$. The measurements of $S_{\rm glass}$ and $S_{\rm conf}$ thus provide us with invaluable information about glass formation approaching $T_{\rm K}$.

The second general route to $S_{\rm conf}$ is to directly enumerate the number of distinct glass states available to the system in equilibrium, $\mathcal{N}$, and use $S_{\rm conf}=\ln \mathcal{N}$. Here, mean-field theory provides a rigorous definition of glass states as free-energy minima. However, just as for ordinary phase transitions (e.g. van der Waals theory) local free-energy minima are no longer infinitely long-lived when physical dimension is finite, and states can no longer be defined precisely. 
Thus, strictly speaking, the complexity that vanishes at $T_K$ in mean-field theory {\it is not defined in finite dimensional systems}. Again, approximations must be performed to measure a physical analog. Two such methods based on the Franz-Parisi free energy~\cite{FP95,FP97} and glassy correlation length,~\cite{BB04} are now available, as discussed in Secs.~\ref{sec:FP} and \ref{sec:PTS}. The existence of an entropy crisis in finite dimension is not directly challenged by the approximate nature of these estimates. To determine whether a Kauzmann transition can occur in finite $d$, one should rather study the effect of finite-dimensional fluctuations within a $d$-dimensional field theory using the Franz-Parisi free-energy as a starting point.\cite{effort1,effort2,effort3,effort4,effort5,effort6} There exists no `proof' that the Kauzmann transition should be destroyed in finite dimensions as divergent conclusions were obtained using distinct approximate field-theoretical treatments. This is a difficult, but pressing, theoretical question for future work.

A popular alternative is the enumeration of potential energy minima using the potential energy landscape (PEL), which was actually proposed long before the development of mean-field theory, first by Goldstein,~\cite{Go69} and further formalized by Stillinger and Weber.~\cite{SW82,stillinger1995topographic} The PEL approach assumes that an equilibrium supercooled liquid resides very close to a minimum of the potential energy, also named inherent structure. Assuming further that each inherent structure corresponds to a distinct glass state, the number of inherent structures,  ${\cal N}_{\rm IS}$, provides a proxy for the configurational entropy, $S_{\rm conf} \approx \ln {\cal N}_{\rm IS}$.  
This assumption offers precise and simple computational methods to estimate the configurational entropy,~\cite{stillinger1988supercooled,SKT99,sastry2001relationship} discussed below in Sec.~\ref{sec:PEL}. 

The \res{identification between inherent structures and the free-energy minima entering the mean-field theory should not be made, as explicit examples were proposed to show that it is generally incorrect.~\cite{BM00,OB17}} Physically, it is believed that free-energy minima may contain a large number of inherent structures. The concept of `metabasins'~\cite{heuer2008exploring} has been empirically introduced to capture this idea, but there is no available method to enumerate the number of metabasins to obtain a configurational entropy. The hard sphere model is a striking example of the difference between energy and free energy minima. In large dimensions, hard spheres undergo an entropy crisis, but it does not correspond to a decrease of the number of inherent structures, \res{which are not defined for due to the discontinous nature of the pair potential.}

Using the PEL approach, several arguments were given to question the existence of a Kauzmann transition in supercooled liquids. By considering localized excitations above inherent structures, Stillinger provided a physical argument showing that the PEL approximation to the configurational entropy can not vanish at a finite temperature.~\cite{stillinger1988supercooled} The effect of such excitations on the free-energy landscape has not been studied, and so this argument does not straightforwardly apply to the random first order transition itself. In the same vein, Donev {\it et al.} directly constructed dense hard disk packings of a binary mixture model to suggest that ${\cal N}_{\rm IS}$ cannot yield a vanishing configurational entropy. This again does not question the Kauzmann transition of that system, since it should be demonstrated that the equilibrium free-energy landscape is sensitive to these artifical inherent states, whose relevance to the equilibrium supercooled fluid is not established.\footnote{In addition, the constructed dense packings are largely demixed and partially crystallized, and it is unclear that these states are relevant for the (metastable) fluid branch.}
Finally, the ambiguous nature of inherent structures becomes obvious when considering colloidal systems composed of a continuous distribution of particle sizes. Starting from a given inherent structure, each permutation of the particle identity provides a different energy minimum and a naive enumeration of the configurational entropy~\cite{stillinger1999exponential} would contain a divergent mixing entropy contribution, again incorrectly suggesting the absence of a Kauzmann transition.~\cite{OB17} A similar argument was proposed for a binary mixture.~\cite{DTS06} The problem of the mixing entropy in the PEL approach is considered further, and solved, in Sec.~\ref{sec:mixing}.

\section{Computer simulations of glass-forming liquids}
\label{sec:computer_simulation}

\subsection{Why perform computer simulations to measure the configurational entropy?}

%- 1985: early studies
%- 1995: MCT Reichmann.
%- 2000: Aging, effective temperature, rheology
%- 2010: Static lengthscales, Kauzmann transition, config
%- 2017: swap progress. 

Let us start with some major steps in computer simulations of supercooled liquids, referring to broader reviews for a more extensive perspective.~\cite{kob99review,kobhouches} Early computational studies date back to the mid-1980s,~\cite{yip,lewis1991atomic,bernu1987soft,roux1989dynamical,wahnstrom1991molecular} followed by intensive works strongly coupled to the development of mode-coupling theory during the 1990s.~\cite{KA95a} The nonequilibrium aging dynamics of glasses,~\cite{kob1997aging} along with concepts of effective temperatures,~\cite{teff1,teff2,teff3} rheology~\cite{yamamoto,berthier2002nonequilibrium} and dynamical heterogeneities~\cite{kob1997dynamical,yamamoto,widmer2004reproducible,BBBCS11} were in the spotlight at the end of the 20th century. The search for a growing static lengthscale,~\cite{BBCGV08} linked to a Kauzmann transition and configurational entropy,~\cite{SKT99,sastry2001relationship} have continuously animated the field until today. Over this period spanning about 3 decades, the numerically-accessible time window increased about as many orders of magnitude, mainly due to improvements in computer hardware. Until 2016, computer studies lagged well behind experiments in terms of equilibrium configurational entropy measurements, 
\res{but recent developments in computer algorithms have been able
to generate, for highly polydisperse systems, equilibrium configuration comparable to experimental data.~\cite{NBC17}}
%but recent developments in computer algorithms have suddenly closed the gap with experiments.
\res{For these models,} temperatures {\it below} the experimental glass transition are now numerically accessible in equilibrium conditions, making computer simulations an essential tool for configurational entropy studies in supercooled liquids.~\cite{ceiling17,berthier2018zero} 

As illustrated in Fig.~\ref{fig:cartoon}, theories for the glass transition need to make predictions for complex observables that reflect nontrivial changes in the supercooled liquid, such as multi-point time correlation functions,~\cite{doliwa2000cooperativity} point-to-set correlations,~\cite{BBCGV08,BK12} non-linear susceptibilities,~\cite{donati2002theory,BBBCEHLP05} as well as properties of the potential and free-energy landscapes.~\cite{sastry1998signatures,wales2003energy,sciortinoPEL} Most of these quantities are extremely challenging, or sometimes even impossible, to measure in experiments. Computer simulations are particularly suitable  because they generate equilibrium density profiles from which any observable can be computed. Obtaining the same information in experiments is possible to some extent in colloidal glasses, but still a challenge in atomistic or molecular glasses. 

Computer simulations take place under perfectly controlled conditions, and are therefore easier to interpret than experiments. All settings are well-defined: microscopic model, algorithm for the dynamics, statistical ensemble (isobaric or isochoric conditions), external parameters, etc. Computer simulations are also very flexible. Since the mean-field theory for the glass transition provides exact predictions for the configurational entropy in infinite dimensions, it is crucial to understand how finite-dimensional fluctuations affect them. Along with current efforts that strive to develop renormalization group approaches to this problem, numerical simulations give precious insights into the effect of dimensionality on the physics of glass formation. Numerical simulations can be performed in any physical dimensions, and the range $d=1 - 12$ was explored in that context.~\cite{yamchi2015inherent,sengupta2012adam,eaves2009spatial,CIPZ11} Even the space topology can be varied.~\cite{STV08,turci2017glass} One can study the effect of freezing a subset of particles with arbitrary geometries by means of computer simulations.~\cite{kim2003effects,BK12,KB13,sandalo,ozawa2015equilibrium} The size of the system under study can be tuned and finite-size scaling analysis can reveal important lengthscales for the glass problem.~\cite{berthier2012finite,karmakar2012finite} 

\subsection{Simple models for supercooled liquids}

% Models
The features associated with the glass transition, such as a dramatic dynamical slowdown and dynamic heterogeneities, are observed in a wide variety of glassy materials composed of atoms, molecules, metallic compounds, colloids and polymers. It may be useful to focus on simple models exhibiting glassy behavior to understand universal features of the glass transition. We consider classical point-like particles with no internal degrees of freedom that interact via isotropic pair potentials. These models may not capture all detailed aspects of glass formation, e.g. $\beta$-relaxations due to slow intramolecular motion in molecules, but their configurational entropy can nevertheless be measured. The numerical study of simple models is especially relevant in the context of configurational entropy, since mean-field theory was precisely derived for such simple models, which allow direct comparison between theory and simulations.~\cite{CKPUZ16} 

\res{For this perspective, we use results for three simple glass-formers to illustrate generic features of entropy measurements.} The Lennard-Jones (LJ) potential was first introduced to model the interaction between neutral atoms and molecules. The interaction potential between two particles separated by a distance $r$ reads
\be
v_{\text{LJ}}(r) = 4 \varepsilon \bigg[  \Big( \frac{\sigma}{r} \Big)^{12} - \Big( \frac{\sigma}{r} \Big)^6    \bigg]~,
\label{eq:LJ}
\ee
where $\varepsilon$ and $\sigma$ set the energy and length scales. % Variations of the LJ potential are commonly used.
The stiffness of the repulsion in the soft-sphere (SS) potential $v_{\text{SS}}(r) = \varepsilon (\sigma/r)^{\nu}$ can be tuned with $\nu$.~\cite{NBC17} The hard-sphere (HS) potential, defined as $v_{\text{HS}}(r) = \infty$ if $r <\sigma$, and $v_{\text{HS}}(r) = 0$ otherwise, models hard-core repulsion between particles of diameter $\sigma$. This highly-idealized model efficiently captures the glass transition phenomenology.\cite{BW09,BEPPSBC08} \res{We recall that for hard spheres, pressure and temperature are no longer independent control parameters, but enter together in the adimensional pressure, $p = P / (\rho T)$, so that $1/p$ replaces the temperature for that system,~\cite{BW09} and directly controls the packing fraction $\phi$ via the equation of state, $\phi = \phi(p)$.}
%In such a model, the energy is zero, and temperature plays only the role of imposing the timescale. The only state parameter is pressure $P$, or equivalently packing fraction $\phi$. 

The homogeneous supercooled liquid is metastable with respect to the crystal in the temperature regime where the configurational entropy is measured, \res{and so the expression ``equilibrium supercooled liquid'' represents, strictly speaking, an abuse of language.} Designing glass-forming models in which crystallization is frustrated, and defining strict protocols to detect crystallization is crucial.\cite{NBC17} Mixtures of different species are good experimental glass-formers: colloidal glasses are made of polydisperse suspensions,~\cite{hunter2012physics} and metallic glasses are alloys of atoms with different sizes.~\cite{chen2011brief} Inspired by experiments, numerical models use particles of different species which differ by their size $\sigma$ or interaction $\varepsilon$. The Kob-Andersen (KA) model is a bidisperse mixture with 80\% larger particles and 20\% smaller particles, interacting via the LJ potential with adjusted parameters $\sigma$ and $\varepsilon$ to describe amorphous ${\rm Ni}_{80} {\rm P}_{20}$ metallic alloys.\cite{KA95a} Many numerical models with good glass-forming ability have been developed,~\cite{bernu1987soft,wahnstrom1991molecular,KA95a,coslovich2009dynamics,Gutierrez:2015,NBC17} although development in computational power now leads to crystallization for some of those models.~\cite{toxvaerd2009stability,NBC17,coslovich2018local,coslovich2018dynamic} Thus, developing new models robust against crystallization is an important research goal. 

While the situation may seem satisfactory to theorists, numerical glass-formers are probably too simplistic for many experimentalists. \res{A wide variety of more realistic glass forming models have been developed and studied~\cite{mossa2002dynamics,giovambattista2004glass,sastry2003liquid,saika2001fragile}. Future developments should aim at designing minimal models for more complex systems and powerful algorithms for efficient simulations, in order to also close this conceptual gap.}

\subsection{Molecular dynamics simulations}

%-how to do a simulation, thermalization then measurements.
The two main classical methods used to simulate the above models are Monte Carlo (MC) and Molecular Dynamics (MD) simulations.\cite{AT89,LandauBinder05} Quantum effects, partially included in \textit{ab initio} simulations, are irrelevant in the present context. 

The course of a numerical simulation is very similar to an experiment. A sample consisting of $N$ particles is prepared and equilibrated (using either MC or MD dynamics) at the desired state point, until its properties no longer change with time. After equilibration is achieved, the measurement run is performed. Common problems are just as in experiments: the sample is not equilibrated correctly, the measurement is too short, the sample undergoes an irreversible change during the measurement, etc.

A noticeable difference between computer and experimental supercooled liquid samples is their size. Numerical studies of the configurational entropy are limited to around $10^4$ particles, to be compared to around $10^{23}$ atoms or molecules in experimental samples. Periodic boundary conditions are applied to the simulation box in order to avoid important boundary effects, and simulate the bulk behavior of `infinitely' large samples. Lengthscales larger than the system size are numerically inaccessible. \res{Up to now, this limit was not really problematic since all relevant lengthscales associated with the glass transition are not growing to impossibly large values, in particular for static quantities. Analysis of dynamic heterogeneity have shown that systems larger than $10^4$ particles are sometimes needed.~\cite{BBBCS11,berthier2012finite,karmakar2014growing}} 

The difference between MD and MC is the way the system explores phase space. The molecular dynamics method simulates the physical motion of $N$ interacting particles. As an input, one defines a density profile $ {\bf r}_0^N$, particle velocities $ {\bf v}_0^N$, and an interaction potential between particles. The method solves the classical equations of motion step by step using a finite difference approach. As an output, one obtains physical particle trajectories $({\bf r}^N(t), {\bf v}^N(t))$ from which thermodynamic quantities can be computed, see Sec.~\ref{section:fromconf2thermo}. By construction, the trajectories sample the microcanonical ensemble. Other ensembles can be simulated by adding degrees of freedom which simulate baths which generate equilibrium fluctuations in any statistical ensemble.\cite{Nose84,Hoover85,AT89}

\begin{figure}
\includegraphics[width=0.95\columnwidth]{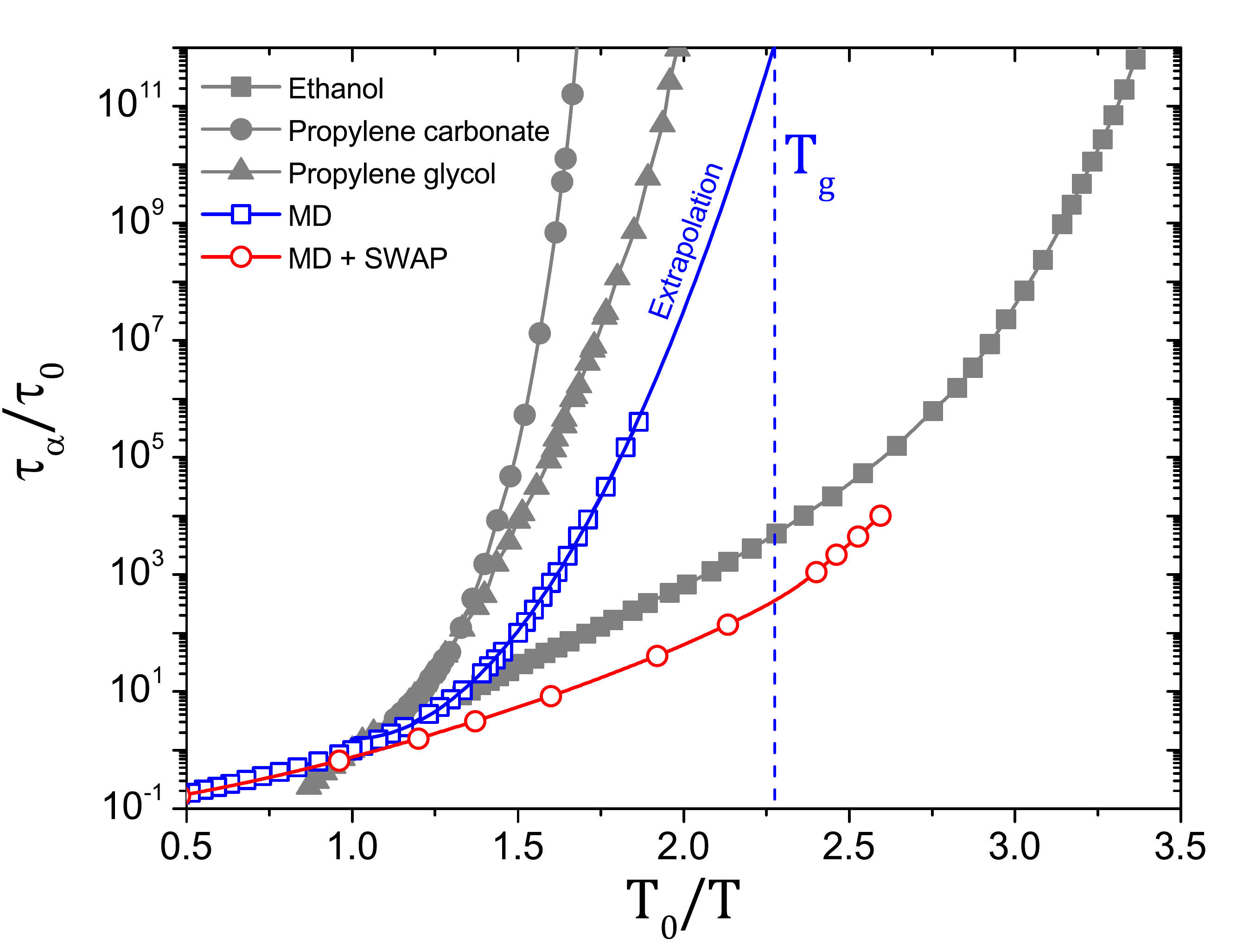}
\caption{Isobaric relaxation time of supercooled liquids as a function of the inverse temperature for ethanol,~\cite{ethanolLunk} propylene carbonate,~\cite{schneider1999broadband} and propylene glycol,~\cite{lunkenheimer2005glassy} as well as the standard molecular dynamics (open squares) and its combination with the swap Monte-Carlo algorithm (open circles)~\cite{berthier2018efficient} for three-dimensional polydisperse soft spheres~\cite{NBC17}. We renormalize axis using the onset of glassy dynamics, ($\tau_0 = 10^{-10}$s in experiments), and the corresponding $T_0$. We fit MD results with a parabolic fit, which provides a reasonable estimate of $T_g$ for this system (vertical dashed line). The SWAP algorithm (open circles) can equilibrate the numerical model well below that $T_{\rm g}$ value.}
\label{fig:swap}
\end{figure}  

Molecular dynamics mimics the physical motion of particles, very much as it  takes place in experiments, but computers are much less efficient than Nature. Long MD simulations of a simple glass model (about a month) can only track the first 4-5 orders of magnitude of dynamical slowdown in supercooled liquids approaching the glass transition, to be compared to 12-13 orders of magnitude in real molecular liquids. In Fig.~\ref{fig:swap}, we show relaxation time $\tau_{\alpha}$ of some molecular liquids of various fragilities,~\cite{angell1995formation,ethanolLunk,schneider1999broadband,lunkenheimer2005glassy} and MD simulations of polydisperse soft spheres under isobaric condition. The temperature range accessible with MD simulations is far from the experimentally relevant regime, and stops well before $T_{\rm g}$ is reached (estimated from a parabolic fit~\cite{elmatad2009corresponding}). 

\res{Recently, efficient software packages for MD have been developed that use the power of graphic cards.~\cite{bailey2017rumd,glaser2015strong} They typically yield a speed-up of about two orders of magnitude over normal MD, which is sufficient to get below the mode-coupling crossover, and thus access interesting new physics and dynamics.~\cite{bailey2017rumd,coslovich2018dynamic}}

\subsection{Beating the timescale problem: Monte Carlo simulations}

%-stochastic Markov process that efficiently samples the equilibrium distribution. Theory, detailed balance. 
Monte Carlo simulations aim at efficiently sampling the configurational space with Boltzmann statistics.\cite{Met53,NB99} A stochastic Markov process is generated in which a given configuration ${\bf r}^N$ is visited with a probability proportional to the Boltzmann factor $e^{-\beta U ({\bf r}^N)}$, where $\beta=1/T$ and $U$ are the inverse of the temperature and the potential energy, respectively. The method only considers configurational, and not kinetic, degrees of freedom, and is suitable for configurational entropy measurements.

A Markov process is defined by the transition probability $T({\bf r}^N \rightarrow {\bf r'}^N)$ to go from configurations ${\bf r}^N$ to ${\bf r'}^N$. To sample configurations with a probability $P({\bf r}^N)$ given by the Boltzmann factor, the global balance condition should be verified
\be
\sum_{{\bf r'}^N} P({\bf r}^N) T({\bf r}^N \rightarrow {\bf r'}^N) = \sum_{{\bf r'}^N} P({\bf r'}^N) T({\bf r'}^N \rightarrow {\bf r}^N)~~.
\label{eq:GB}
\ee
We consider a stronger condition and impose the equality in Eq.~(\ref{eq:GB}) to be valid for each new state ${\bf r'}^N$. This detailed balance condition reads
\be
\frac{T({\bf r}^N \rightarrow {\bf r'}^N) }{T({\bf r'}^N \rightarrow {\bf r}^N)} = \frac{P({\bf r'}^N)}{P({\bf r}^N)} = \exp \left[-\beta \left(U ({\bf r'}^N) - U ({\bf r}^N) \right)\right]~~.
\label{eq:DB}
\ee
In practice, $T({\bf r}^N \rightarrow {\bf r'}^N)  = \alpha({\bf r}^N \rightarrow {\bf r'}^N) \times \mathsf{acc}({\bf r}^N \rightarrow {\bf r'}^N)$, where $\alpha$ and $\mathsf{acc}$ are the probabilities to propose a trial move and to accept it, respectively. We consider a symmetric matrix $\alpha$ for trials such that the matrix $\mathsf{acc}$ obeys the same equation as $T$ in Eq.~(\ref{eq:DB}). If trial moves are accepted with probability $\mathsf{acc}({\bf r}^N \rightarrow {\bf r'}^N) = \min \left\{1, \exp \left[- \beta (U ({\bf r'}^N) - U ({\bf r}^N) ) \right] \right\}$ (Metropolis criterion),~\cite{Met53} the configurations are drawn from the canonical distribution at equilibrium at the desired temperature. 

Contrary to MD simulations, dynamics in a Monte Carlo simulation is not physical, since it results from a random exploration of configurational space. This is actually good news, since there is a considerable freedom in the choice of trial moves, opening the possibility to beat the numerical timescale problem illustrated in Fig.~\ref{fig:swap}. The choice of trial move depends on the purpose of the numerical simulation. A standard trial move consists in selecting a particle at random and slightly displacing it. For small steps, the dynamics obviously resembles the (very physical) Brownian dynamics.~\cite{BK07} 

\res{
Efficient Monte Carlo simulations should in principle be possible using lattice models for glasses, which would use discrete rather than continuous degrees of freedom. This approach has been heavily used to analyse models based on dynamic facilitation such as kinetic Ising models~\cite{PhysRevLett.53.1244}, or plaquette models~\cite{JBG05} but the entropy does not play any central role in these models. Lattice glass models were introduced as lattice models that have, in some controlled mean-field limit, a random first order transition,~\cite{BM01,RBMM04} but simulation studies of finite dimensional versions of these models remain scarce,~\cite{DRB10} and we are aware of no study of configurational entropy in such lattice models. 
}

If instead efficient equilibration is targeted, more efficient but less physical trial moves should be preferred. In the SWAP algorithm,~\cite{tsai1978structure,gazzillo1989equation,frenkel2001understanding,GP01,fernandez2007phase,Gutierrez:2015,NBC17} trial moves combine standard displacement moves and attempts to swap the diameters of two randomly chosen particles. Since the trial moves satisfy detailed balance in Eq.~(\ref{eq:DB}), the SWAP algorithm by construction generates equilibrium configurations from the canonical distribution.

Using continuously polydisperse samples, this algorithm outperforms standard MC or MD, as equilibrium liquids can be generated at temperatures below the experimental glass transition.~\cite{NBC17} In Fig.~\ref{fig:swap}, we show the equilibrium relaxation time $\tau_{\alpha}$ of an Hybrid scheme of MD and SWAP MC developed recently in Ref.~\onlinecite{berthier2018efficient}. The relaxation dynamics with this scheme is significantly faster than with standard MD, which makes equilibration of the system possible even below the estimated experimental glass transition temperature $T_{\rm g}$. Accessing numerically these low temperatures is crucial to compare simulations and experiments. From a theoretical perspective, the concept of metastable state applies far better at low temperatures. In particular, numerical estimates for the configurational entropy become more meaningful in these extreme temperature conditions. 

To conclude, Monte Carlo simulations are very relevant in the present context, because their flexibility allows us to compute and compare different estimates for the configurational entropy of supercooled liquids.~\cite{ceiling17,ozawa2018configurational} These measurements are done under perfectly controlled conditions, in a temperature regime relevant to experimental works, and even at lower temperatures.~\cite{berthier2018zero}

%\MO{Huge acceration of SWAP provides not only practical application for obtaining low temperature equilibrium configurations, but also deep scientific question about the mechanism of glassy slow dynamics. Especially the role of the static correlation length on the slow dynamics has been discussed.}

\subsection{From microscopic configurations to observables}

\label{section:fromconf2thermo}

The output of a numerical simulation consists in a series of equilibrium configurations. To measure an observable numerically, one must first express it as a function of the positions of the particles.

Static quantities describing the structure of the liquid are easily computed.\cite{Hansen} In particular, the density field is given by 
\be
\rho({\bf r} ) = \sum_{i=1}^{N} \delta ({\bf r}  - {\bf r} _i) ~.
\label{eq:densityfield}
\ee

Two-point static density correlation functions such as the pair correlation function 
\be
g({\bf r} ) = \frac{1}{\rho N}   \left\langle \sum_{i \neq j }^{}   \delta({\bf r}  + {\bf r}_i - {\bf r}_j)  \right\rangle  ~,
\ee
where $\rho = N/V$ is the number density and the bracket indicate an ensemble average at thermal equilibrium, and the structure factor 
\be
S({\bf k} ) = \frac{1}{N} \left\langle \rho_{\bf k}  \rho_ {-{\bf k} } \right\rangle ~,
\ee
are evaluated, where $ \rho_{\bf k}  = \sum_{i = 1}^{N} e^ {i {\bf k}  \cdot {\bf r} _i} $ is the Fourier transform of the density field. Even if these quantities are not relevant to describe the dynamical slowdown of the supercooled liquid (see Fig.~\ref{fig:cartoon}), they are convenient to detect instabilities of the homogeneous fluid (crystallization, fractionation). 
Thermodynamic quantities (such as energy, pressure), and their fluctuations (e.g. specific heats, compressibility), related to macroscopic response functions, can be computed directly from the two-point structure of the liquid.

As presented in Sec.~\ref{sec:MF}, the relevant order parameter for the glass transition is the overlap $Q$ that quantifies the similarity of equilibrium density profiles. This quantity compares the coarse-grained density profiles of two configurations, to remove the effect of short-time thermal vibrations. Numerically, the following definition is very efficient
\be
Q = \frac{1}{N} \sum_{i, j} \theta(a-|{\bf r}_{1 i}-{\bf r}_{2 j}|)~~,
\label{eq:overlap}
\ee
where ${\bf r}_1$ and ${\bf r}_2$ are the positions of particles in distinct configurations, and $\theta(x)$ is the Heaviside step function. The parameter $a$ is usually a small fraction (typically $0.2 - 0.3$) of the particle diameter. The overlap is by definition equal to 1 for two identical configurations, \res{it is slightly smaller than 1 due to the effect of vibrations,} and becomes close to zero (more precisely $4 \pi a^3 \rho / 3 \ll 1$) for uncorrelated liquid configurations at density $\rho$.  

\section{Configurational entropy by estimating a `glass' entropy}

\label{sec:PEL}

\subsection{General strategy}

The configurational entropy enumerates the number of distinct glass states. One possible strategy to achieve this enumeration is to first estimate the total number of configurations, or phase space volume, ${\cal N}_{\rm tot}$. If one can then measure the number of configurations belonging to the same glass state, ${\cal N}_{\rm glass}$, the number of glass states ${\cal N}_{\rm conf}$ can be deduced, ${\cal N}_{\rm conf} = {\cal N}_{\rm tot} / {\cal N}_{\rm glass}$. Taking the logarithm of ${\cal N}_{\rm conf}$ yields the configurational entropy %Following this spirit, we estimate
\begin{equation}
S_{\rm conf} = S_{\rm tot} - S_{\rm glass}. 
\end{equation}
Whereas the measurement of the total entropy $S_{\rm tot}$ is straightforward, the art of measuring the configurational entropy lies in the quality of the unavoidable approximation made to determine $S_{\rm glass}$. Recall that experimentalists typically use $S_{\rm glass} \approx S_{\rm xtal}$. This is not a practical method for simulations, because numerical models which can crystallize are generally very poor glass-formers. In this section, we describe several possible strategies to measure $S_{\rm glass}$ which do not rely on the knowledge of the crystal state, and present their limitations. 

Let us now introduce our notations for entropy calculations. We consider a $M$-component system in the canonical ensemble in $d$ spatial dimensions, with $N$, $V$, and $T=1/\beta$ the number of particles, volume, and temperature, respectively. We fix the Boltzmann constant to unity. We take $M = N$ to treat continuously polydisperse systems. The concentration of the $m$-th species is $X_m=N_m/N$, where $N_m$ is the number of particles of the $m$-th species ($N=\sum_{m=1}^M N_m$). A point in position space is denoted as ${\bf r}^N=({\bf r}_1, {\bf r}_2, \cdots, {\bf r}_N)$. For simplicity, we consider equal masses, irrespective of the species, which we set to unity.

For this system, the following partition function in the canonical ensemble is conventionally used~\cite{frenkel2014colloidal}
\begin{equation}
Z = \frac{\Lambda^{-Nd}}{\Pi_{m=1}^M N_m! } \int_V \mathrm{d} {\bf r}^N e^{-\beta U ({\bf r}^N)},
\label{eq:Z_old}
\end{equation}
where $\Lambda$ and $U ({\bf r}^N)$ are respectively the de Broglie thermal wavelength and the potential energy. 
%We set the Planck constant to unity, $\hbar=1$. 
The only fluctuating variables are the configurational degrees of freedom ${\bf r}^N$, since momenta are already traced out in Eq.~(\ref{eq:Z_old}). 

\subsection{Total entropy $S_{\rm tot}$}

An absolute estimate of the total entropy at a given state point can be obtained by performing a thermodynamic integration from a reference point where the entropy is exactly known,~\cite{SKT99,sastry2000evaluation,CPV00,AF07} typically the ideal gas at $\rho \to 0$ or  $\beta \to 0$. 
This approach works for all state points which can be studied in equilibrium conditions, and are connected to the reference point by a series of equilibrium state points. This is usually doable also in most experiments. However, this constraint prevents \res{a direct analysis of the entropy of vapor-deposited ultrastable glasses} produced directly at very low temperature.
In practice, to perform the thermodynamic integration and access $S_{\rm tot}$, we need to distinguish between continuous `soft' interaction potentials, such as the Lennard-Jones potential, and discontinuous `hard' potentials, as in the hard sphere model:
\begin{eqnarray}
S_{\rm tot} &=& S_{\rm id} + \beta E_{\rm pot}(\beta) - \int_0^{\beta} \mathrm{d} \beta' E_{\rm pot}(\beta') \quad {\rm (soft)}, \label{eq:total_soft} \\
S_{\rm tot} &=& S_{\rm id} - N \int_0^{\rho} \mathrm{d} \rho' \frac{(p(\rho')-1)}{\rho'} \quad {\rm (hard)},
\label{eq:total_hard}
\end{eqnarray}
where $S_{\rm id}$, $E_{\rm pot}$ and $p = P / (\rho T)$ are the ideal gas entropy, the averaged potential energy, and the reduced pressure, respectively. The ideal gas entropy $S_{\rm id}$ can be written as
\begin{equation}
S_{\rm id}= N \frac{(d+2)}{2} - N \ln \rho - N \ln \Lambda^d + S_{\rm mix}^{(M)}, 
\label{eq:entropy_ideal_gas}
\end{equation} 
where $S_{\rm mix}^{(M)}$ is the mixing entropy of the ideal gas,
\begin{equation}
S_{\rm mix}^{(M)} = \ln \left( \frac{N!}{\Pi_{m=1}^M N_m!}\right).
\label{eq:s_mix_M}
\end{equation}
When $M$ is finite and $N_m \gg 1$, Stirling's approximation can be used, $\ln N_m! \simeq N_m \ln N_m -N_m$, and Eq.~(\ref{eq:s_mix_M}) reduces to $S_{\rm mix}^{(M)}/N=-\sum_{m=1}^M X_m \ln X_m$.

\begin{figure}
\includegraphics[width=0.95\columnwidth]{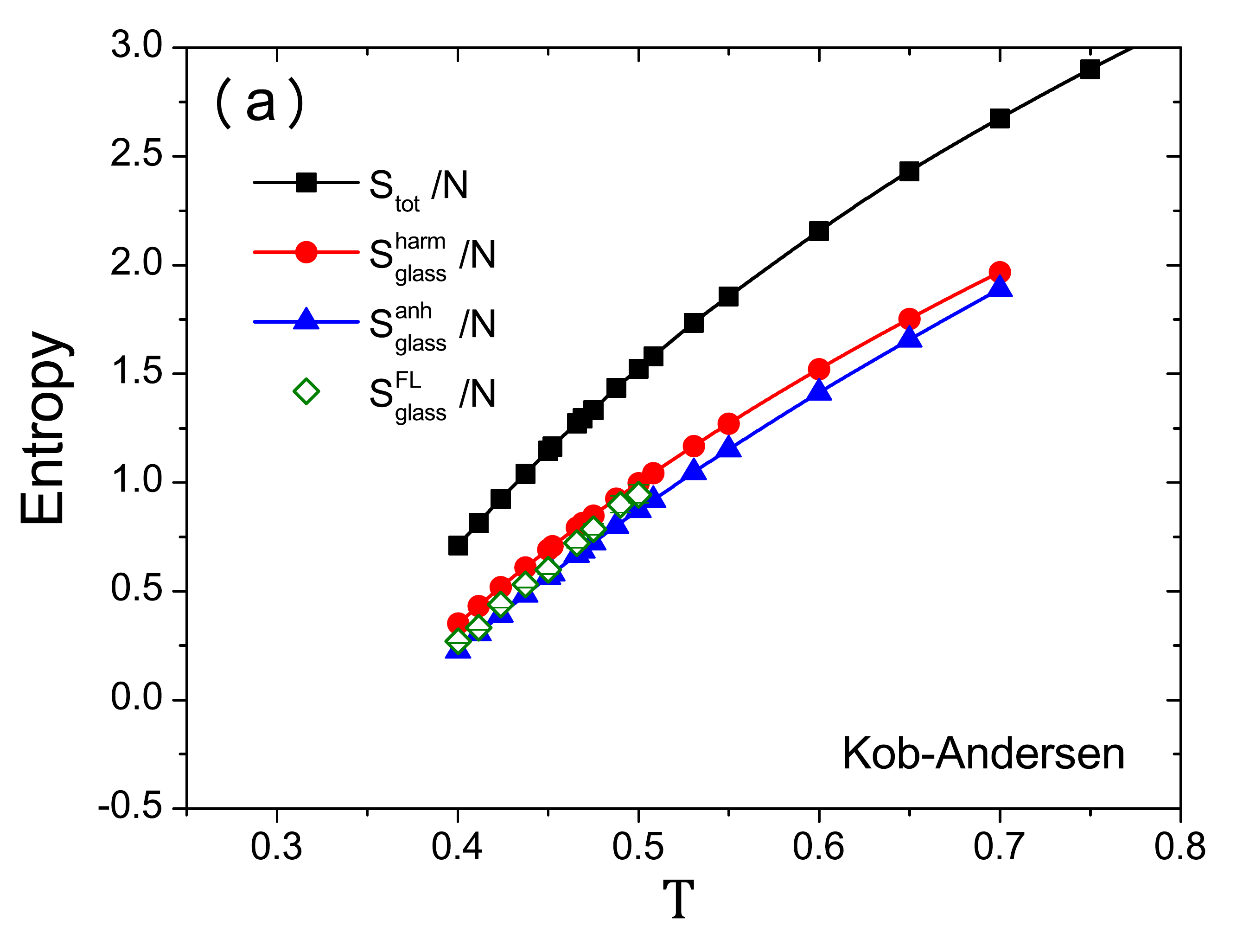}
\includegraphics[width=0.95\columnwidth]{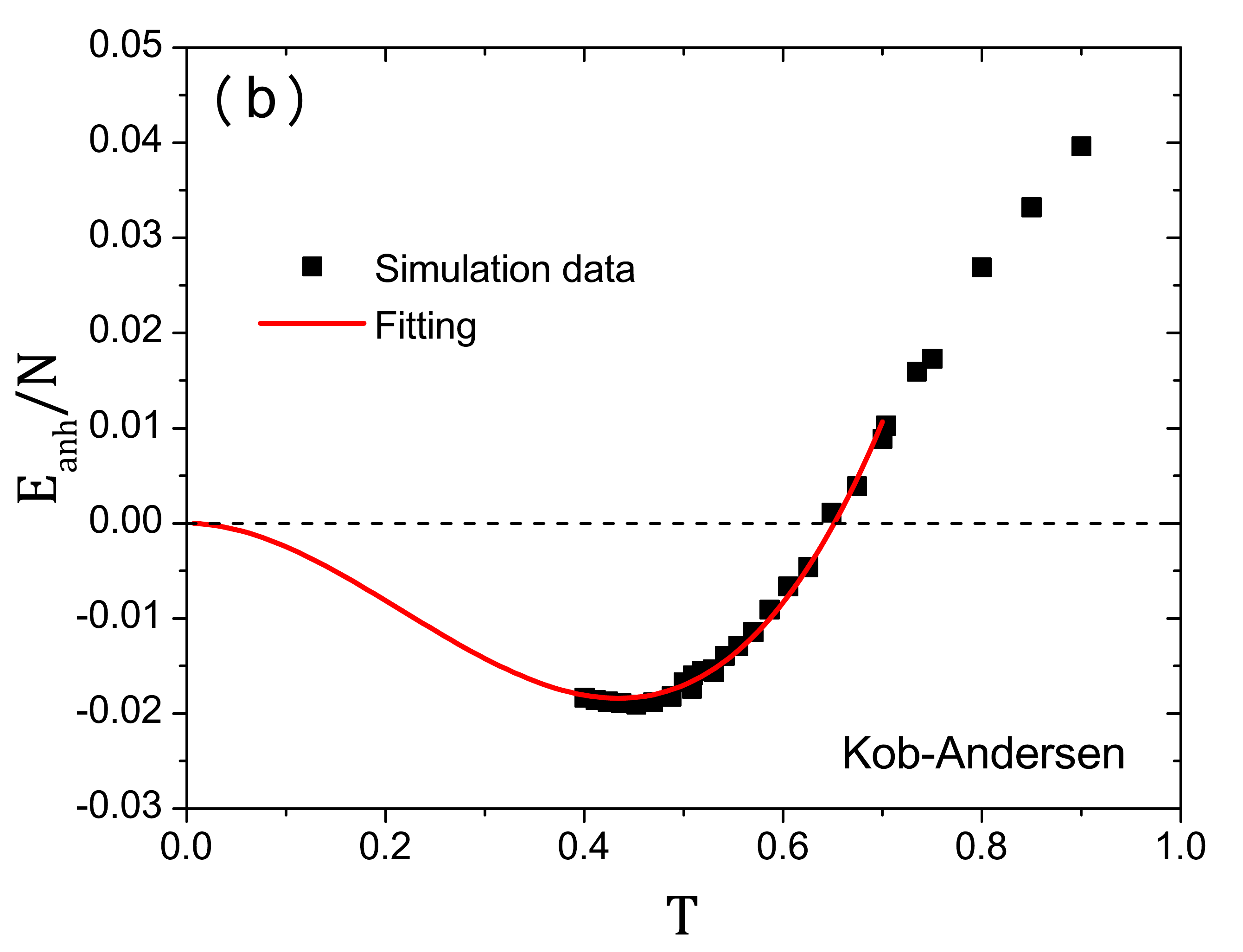}
\includegraphics[width=0.95\columnwidth]{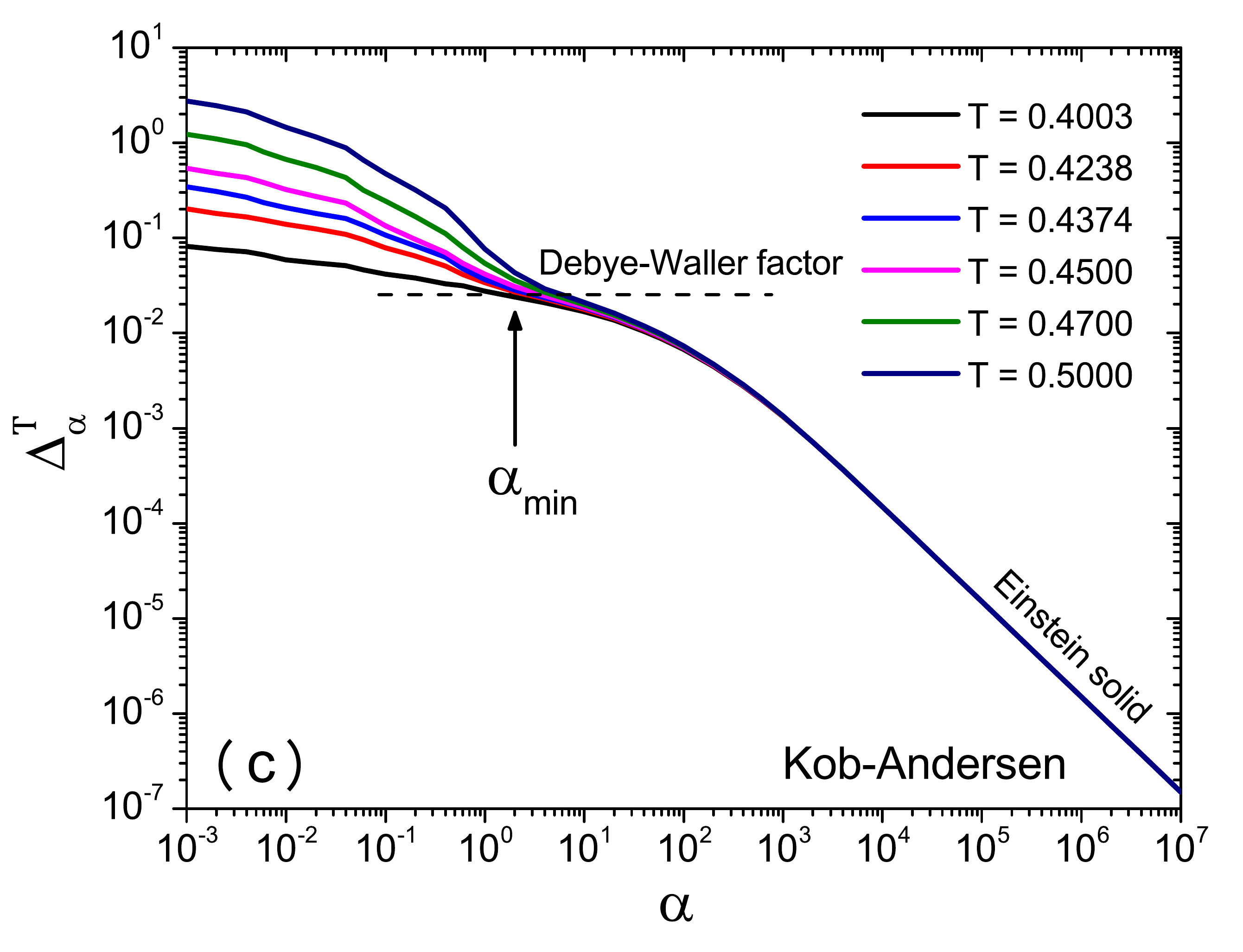}
\caption{(a) Total entropy $S_{\rm tot}$ and various estimates of the glass entropy $S_{\rm glass}$: harmonic $S_{\rm glass}^{\rm harm}$, with anharmonic correction $S_{\rm glass}^{\rm anh}$, and the Frenkel-Ladd entropy $S_{\rm glass}^{\rm FL}$.
(b) Anharmonic energy $E_{\rm anh}$ from simulations (black points) and polynomial fitting to third order (red line). 
(c) Constrained mean-squared displacement in the Frenkel-Ladd method. The dashed horizontal line correspond to the Debye-Waller factor independently measured in the bulk dynamics at the lowest temperature. The vertical arrow indicates $\alpha_{\rm min}$.}
\label{fig:KA}
\end{figure}

As a representative example, Fig.~\ref{fig:KA}(a) shows the temperature dependence of the numerically measured total entropy in the Kob-Andersen model.~\cite{KA95a} It decreases monotonically with decreasing the temperature.

\subsection{Inherent structures as glass states}

The first strategy that we describe to identify glass states and estimate $S_{\rm glass}$ is based on the potential energy landscape (PEL).~\cite{Go69,SW82,stillinger1995topographic,sciortinoPEL,heuer2008exploring}
The central idea is to assume that each configuration can be decomposed as 
\begin{equation}
{\bf r}^N = {\bf r}^N_{\rm IS} + \Delta {\bf r}^N,
\label{eq:decomposition1}
\end{equation} 
where ${\bf r}^N_{\rm IS}$ is the position of the `inherent structure', \textit{i.e.} the potential energy minimum closest to the original configuration. This trivial decomposition becomes meaningful if one makes the central assumption that {\it different inherent structures represent distinct glass states}. It follows immediately that the glass entropy, $S_{\rm glass}$, then quantifies the size of the basin of attraction of inherent structures. 

%Here we consider the inherent structure approach to estimate $S_{\rm glass}$. This approach perform partitioning the configuration space into the basins of attraction of inherent structures, defined as minima of the potential energy landscape, which consists mathmatically precise procedure~\cite{SKT99,sastry2000evaluation,CPV00}. Basic idea is that each inherent structure would correspond to a glass state that we wish to enumerate for $S_{\rm conf}$. Thus the glass entropy is represented by the entropy of the inherent structure.

Assuming that temperature is low, $\Delta {\bf r}^N$ can be treated in the harmonic approximation. Expanding the potential energy $U({\bf r}^N)$ around the inherent structure ${\bf r}_{\rm IS}^N$, one gets 
\begin{equation}
U_{\rm harm}({\bf r}^N) \simeq  U({\bf r}_{\rm IS}^N) + \frac{1}{2} \sum_{i, j} \frac{\partial^2 U({\bf r}_{\rm IS}^N) }{\partial {\bf r}_i \partial {\bf r}_j} \Delta {\bf r}_i \Delta {\bf r}_j. 
\end{equation}
Injecting this expansion in the partition function, Eq.~(\ref{eq:Z_old}), gives 
\begin{equation}
Z_{\rm harm} =  e^{-\beta U({\bf r}_{\rm IS}^N)} \Pi_{a=1}^{Nd} (\beta \hbar \omega_a)^{-1},
\label{eq:Z_harm}
\end{equation}
where $\omega_a^2$ are the eigenvalues of the Hessian matrix. We also considered that each inherent structure is realized $\Pi_{m=1}^M N_m!$ times in the phase space volume, as permuting the particles within each specie leaves the configuration unchanged (see related argument in mean-field theory)~\cite{BCPZ09,ikeda2016note,ikeda2018effect}. This factorial term cancels out with the denominator in Eq.~(\ref{eq:Z_old}).  
%We note that Eq.~(\ref{eq:Z_harm}) does not contain the factrial term, $\Pi_{m=1}^M N_m!$, (which turns into the mixing entropy) seen in Eq.~(\ref{eq:Z_old}). Because there is a factorial degeneracy of the basin specified by ${\bf r}_{\rm IS}^N$ in the phase space volume. In particular, these basins can be generated by permutation of the particles. Thus, we have to take into account the number of physically identical basins to evaluate $Z_{\rm harm}$.
%Here we assume that the degeneracy is simply given by the number of configurations obtained by permuting particles within each species, and this corresponds to $\Pi_{m=1}^M N_m!$, which exactly cancells out the denominator seen in Eq.~(\ref{eq:Z_old}). 

We have implicitely assumed that exchanging two distinct particles produces a different inherent structure,~\cite{stillinger1999exponential} which is consistent with the identification of energy minima as glass states. Physically, this implies that there is no mixing entropy associated with inherent structures. As realized recently,~\cite{OB17} this assumption produces unphysical results for systems with continuous polydispersity. 

Averaging over independent inherent structures (denoted by $\left\langle (\cdots) \right\rangle_{\rm IS}$), the free energy of the harmonic solid is obtained,
\begin{eqnarray}
\label{eq:F_harm}
-\beta F_{\rm harm} &=& \left\langle \ln Z_{\rm harm} \right\rangle_{\rm IS} \\
&=&  \nonumber
 -\beta \left\langle U({\bf r}_{\rm IS}^N) \right\rangle_{\rm IS} - \left\langle \sum_{ a=1}^{Nd} \ln (\beta \hbar \omega_a) \right\rangle_{\rm IS}.
\end{eqnarray}
%where $\left\langle \cdots \right\rangle_{\rm IS}$ is the average over the inherent structures.
The internal energy of the harmonic solid is
\begin{equation}
E_{\rm harm}=\frac{Nd}{2} T + \left\langle U({\bf r}_{\rm IS}^N) \right\rangle_{\rm IS} + \frac{Nd}{2} T,
\label{eq:E_harm}
\end{equation}
where the first and last terms are the kinetic and harmonic potential energies.
Using Eqs.~(\ref{eq:F_harm}) and (\ref{eq:E_harm}), we finally obtain the glass entropy in the harmonic approximation:
\begin{eqnarray}
S_{\rm glass}^{\rm harm} & = & \beta \left( E_{\rm harm} - F_{\rm harm} \right), 
\nonumber \\
&=& \left\langle \sum_{\rm a=1}^{Nd} \left\{ 1- \ln (\beta \hbar \omega_a) \right\} \right\rangle_{\rm IS}.
\label{eq:S_harm}
\end{eqnarray}

In practice, this method requires the production of a large number of independent inherent structures, obtained by performing energy minimizations from equilibrium configurations using widespread algorithms such as the steepest decent or conjugate gradient methods,~\cite{nocedal2006numerical} or FIRE.~\cite{bitzek2006structural} The energy $U({\bf r}_{\rm IS})$ is measured, and the Hessian matrix is diagonalized to get the eigenvalues $\omega_a^2$. Using Eq.~(\ref{eq:S_harm}), these measurements then provide the glass entropy $S_{\rm glass}^{\rm harm}$. 
The numerical results for $S_{\rm glass}^{\rm harm}(T)$ in the Kob-Andersen model are shown in Fig.~\ref{fig:KA}(a).
%$S_{\rm harm}$ as a function of $T$ is shown in Fig.~\ref{fig:KA}(a).
The difference $S_{\rm tot}-S_{\rm glass}^{\rm harm}$ is a widely used practical definition of the configurational entropy in computer simulations.~\cite{SKT99,sastry2000evaluation,sastry2001relationship,flenner2006hybrid,banerjee2014role,ozawa2015equilibrium}

\subsection{Anharmonicity}

Although presumably not the biggest issue, it is possible to relax the harmonic assumption in the above procedure.~\cite{sciortinoPEL}
%The harmonic description of the inherent structure should be a good starting point, but of cause the system is not perfectly harmonic in reality~\cite{sciortinoPEL,heuer2008exploring}.
%Here we discuss how to estimate anharmonic contributions based on the inherent structure approach.
First, the anharmonic energy, $E_{\rm anh}$, is obtained by subtracting the harmonic energy in Eq.~(\ref{eq:E_harm}) from the total one, 
\begin{equation}
E_{\rm anh}  =  E_{\rm pot} - \left\langle U({\bf r}_{\rm IS}^N) \right\rangle_{\rm IS} - \frac{Nd}{2} T.
\end{equation}
%Here we evaluate the entropy $S_{\rm anh}$ associated with the observed anhrmonicity. Assuming that the harmonic description is peferectly valid at $T=0$ (hence $S_{\rm anh}(T=0)=0$), one can estimate $S_{\rm anh}(T)$ by a thermodynamic integration of the specific heat ``inside'' a given inherent structure, from $T'=0$ to $T'=T$. This is computationally difficult since finte temperature makes the system escape from the given inherent structure, especially at higher temperature~\cite{sciortinoPEL}. 
The anharmonic contribution to the entropy can then be estimated as 
%Instead one can approximately evaluate $S_{\rm anh}(T)$ by the temperature evolution of $E_{\rm anh}$ by
\begin{equation}
S_{\rm anh} = \int_0^T  \frac{dT'}{T'} \frac{\partial E_{\rm anh}(T')}{\partial T'},
\label{eq:S_anh}
\end{equation}
%We would expect that this approximate way captures the sign and order of true anhamonic entropy~\cite{sciortinoPEL}.
which requires a low-temperature extrapolation of the measured $E_{\rm anh}(T)$. This can be done using an empirical polynomial fitting, $E_{\rm anh}(T)=\sum_{k \geq 2} a_k  T^k$, where the sum starts at $k=2$ to ensure a vanishing anharmonic specific heat at $T=0$. By substituting this expansion in Eq.~(\ref{eq:S_anh}), we obtain
\begin{equation}
S_{\rm anh}(T) = \sum_{k \geq 2} \frac{k}{k-1} a_k T^{k-1}.
\label{eq:S_anh2}
\end{equation}

We show the numerically measured $E_{\rm anh}$ for the Kob-Andersen model, along with its polynomial fit in Fig.~\ref{fig:KA}(b). The non-trivial behavior of $E_{\rm anh}$ suggests that the harmonic description overestimates phase space at low $T$, but underestimates it at high $T$, a trend widely observed across other fragile glass-formers.~\cite{mossa2002dynamics,sciortinoPEL} The resulting 
$S_{\rm anh}$ using Eq.~(\ref{eq:S_anh2}) is thus negative and is of the order of $S_{\rm anh}/N \approx -0.1$, which is a small but measurable correction to $S_{\rm conf}$.
As a result, the improved glass entropy $S_{\rm glass}^{\rm anh} = S_{\rm glass}^{\rm harm} + S_{\rm anh}$ is slightly smaller than the harmonic estimate, as shown in Fig.~\ref{fig:KA}(a).

\subsection{Glass entropy without inherent structures}

\label{sec:FL}

\res{The identification of inherent structures with glass states is a strong assumption which can be explicitly proven wrong in some model systems.~\cite{BM00,OB17,ozawa2018ideal}} Moreover, inherent structures cannot be defined in the hard sphere model \res{(because minima of the potential energy cannot be defined),} which is obviously an important theoretical model to study the glass transition. 
%Because the inherent structures of hard spheres correspond to jammed packing~\cite{MKK08,CBS09,ozawa2012jamming}, which is not suitable to measure the glass entropy.

A more direct route to a glass entropy which automatically includes all anharmonic contributions and can be used for hard spheres is obtained by using the following decomposition, ~\cite{frenkel1984new,coluzzi1999thermodynamics,sastry2000evaluation,foffi2005,AF07,ceiling17,williams2018experimental,ozawa2018ideal}
\begin{equation}
{\bf r}^N = {\bf r}_{\rm ref}^N + \delta {\bf r}^N,
\label{eq:decomposition2}
\end{equation} 
where ${\bf r}_{\rm ref}^N$ is a reference equilibrium configuration. The first difference with Eq.~(\ref{eq:decomposition1}) is that inherent structures do not appear, since deviations are now measured from a given equilibrium configuration.

The second difference is the strategy to estimate the size of the basin surrounding ${\bf r}_{\rm ref}^N$, which makes use of a constrained thermodynamics integration about the fluctuating variables $\delta {\bf r}^N$.
%@@ sketch of 'basins'? or simply a table of sketches at the end? @@
%Since we identify the presence of anharmonicity, we wish to estimate the anharmonic contributions more precisely.Here we compute the vibrational entropy by the Frenkel-Ladd method~\cite{frenkel1984new} that is now widely used in glass physics~\cite{frenkel1984new,coluzzi1999thermodynamics,sastry2000evaluation,foffi2005,AF07,ceiling17,williams2018experimental,ozawa2018ideal}.
%In short, the Frenkel-Ladd method utilizes a thermodynamic integration from the Einstein solid reference state, using an external harmonic potential as a perturbation term.
The potential energy of the system is 
%We denote the potential energy of the target system by 
$\beta U({\bf r}^N)$, is augmented by a harmonic potential to constrain $\delta {\bf r}^N$ to remain small, leading to  
%
%reside close to the reference configuration so that the total energy reads
%A harmonic constraint with spring constant $\alpha$ on the target system $\beta U_0({\bf r}^N)$ as described by
\begin{equation}
\beta U_{\alpha}({\bf r}^N, {\bf r}_{\rm ref}^N) = \beta U({\bf r}^N) + \alpha \sum_{i=1}^{N} | {\bf r}_i - {\bf r}_{{\rm ref}, i} |^2.
\label{eq:Hamiltonian_old}
\end{equation}
%where ${\bf r}_0^N$ is a reference equilibrium configuration drawn from the Boltzmann distribution of the target system. 
%In this approach, ${\bf r}_0^N$ is a randomly chosen equilibrium configuration of the liquid~\cite{AF07,foffi2005,coluzzi1999thermodynamics}, so that the Frenkel-Ladd method implicitly assumes that the vibrational entropy associated with any reference configuration belonging to a given basin is the same for all configurations of that basin. Thus this approach does not reply on quanties computed from the inherent structure.

We consider the statistical mechanics of a given basin, specified by ${\bf r}_{\rm ref}^N$, under the harmonic constraint. The partition function and the corresponding statistical average are
\begin{eqnarray}
Z_{\alpha} &=&  \Lambda^{-Nd} \int_V \mathrm{d} {\bf r}^N  e^{-\beta U_{\alpha}({\bf r}^N, {\bf r}_{\rm ref}^N)}, \label{eq:zalpha} \\
\left\langle (\cdots) \right\rangle_{\alpha}^{\rm T} &=& \frac{\int_V \mathrm{d} {\bf r}^N  (\cdots) e^{ - \beta U_{\alpha}({\bf r}^N, {\bf r}_{\rm ref}^N)}   }{\int_V \mathrm{d} {\bf r}^N  e^{ - \beta U_{\alpha}({\bf r}^N, {\bf r}_{\rm ref}^N)}  }. \label{eq:T} 
\end{eqnarray}
Note that the factorial term $\Pi_{m=1}^M N_m!$ in Eq.~(\ref{eq:zalpha}) is treated as in Eq.~(\ref{eq:Z_harm}) within the PEL approach. 
%this approach e conventional Frenkel-Ladd method does not include the factrial term and the mixing entropy due to the same reason as Eq.~(\ref{eq:Z_harm}).
%Note also that the subscript T in Eq.~(\ref{eq:T}) represents that this average can be numerically evaluated by the normal MC with the translational displacements~\cite{frenkel2001understanding}.
We consider the entropy of a constrained system as $S_{\alpha}=\beta (E_{\alpha}-F_{\alpha})$, where $\beta E_{\alpha} = \frac{Nd}{2} + \beta \left\langle U_{\alpha}({\bf r}^N, {\bf r}_{\rm ref}^N) \right\rangle_{\alpha}^{\rm T}$ and $\beta F_{\alpha}=- \ln Z_{\alpha}$ are the internal energy and free energy, respectively. 

In the glass phase, the system remains close to the reference configuration for any value of $\alpha$, including $\alpha=0$. For the liquid, this is true only for times smaller than the structural relaxation time. For $\alpha$ small but finite, however, the system must remain close to the reference configuration and explore the basin whose size we wish to estimate. We therefore define the glass entropy in the Frenkel-Ladd method as~\cite{frenkel1984new} 
\begin{equation}
S_{\rm glass}^{\rm FL} = \lim_{\alpha_{\rm min} \to 0} \overline{S_{\alpha_{\rm min}}},
\label{eq:S_FL_def}
\end{equation}
where $\overline{(\cdots)}$ represents an average over the reference configuration. The limit in Eq.~(\ref{eq:S_FL_def}) is a central approximation in this method, which is directly related to conceptual problems summarised in Sec.~\ref{sec:conceptual}. Because metastable glass states are not infinitely long-lived in finite dimensions, a finite value of $\alpha$ is needed to prevent an ergodic exploration of the configuration space, and the limit in Eq.~(\ref{eq:S_FL_def}) is difficult to take in practice. 
%This is the main approximation performed in this approach. 
The choice of $\alpha_{\rm min}$  amounts to defining `by hand' the glass state as the configurations that can be reached at equilibrium for a spring constant $\alpha_{\rm min}$. 

%crucial both conceptually and practically. Although the naive limit leads back to the liquid state, here we wish to compute the entropy of a basin characterised by a finite lifetime. To this end, we need to keep $\alpha_{\rm min}$ finite, to prevent the exploration of a different basin during the thermodynamic integration, and we instead make a simple extrapolation of $\alpha_{\rm min}$ from a finite $\alpha_{\rm min}$ value where a basin is well-defined, down to zero. This kind of extrapolation is inevitable in handling basin in finite dimensions, which all have a finite lifetime. Our practical solution to accurately perform the limit is explained below. 

The practical details are as follows. 
At very large $\alpha$ $(=\alpha_{\rm max})$, the entropy is known exactly because the second term in the right hand side of Eq.~(\ref{eq:Hamiltonian_old}) is dominant. The entropy of the system is described by the Einstein solid,
\begin{equation}
S_{\alpha_{\rm max}}=\frac{Nd}{2} - N \ln \Lambda^d - \frac{Nd}{2} \ln \left( \frac{{\alpha_{\rm max}}}{\pi} \right).
\end{equation}
By performing a thermodynamic integration from $\alpha_{\rm max}$, one gets $S_{\alpha_{\rm min}}$, and thus $S_{\rm glass}^{\rm FL}$ from Eq.~(\ref{eq:S_FL_def}) 
\begin{equation}
S_{\rm glass}^{\rm FL}
= S_{\alpha_{\rm max}} 
+ N \lim_{\alpha_{\rm min} \to 0}  \int_{\alpha_{\rm min}}^{\alpha_{\rm max}} \mathrm{d} \alpha \Delta_{\alpha}^{\rm T},
\label{eq:S_FL}
\end{equation}
%Eq.~(\ref{eq:S_FL_def}) can be derived by a thermodynamic integration of the mean-squared displacement $\Delta_{\alpha}^{\rm T}$ over $\alpha$ from the Einstein solid limit ($\alpha=\alpha_{\rm max}$) to small $\alpha=\alpha_{\rm min}$ regimes, expressed as
%\begin{eqnarray}
%S_{\rm FL} &=& S_{\rm Ein}(\alpha_{\rm max})  + N \lim_{\alpha_{\rm min} \to 0} \int_{\alpha_{\rm min}}^{\alpha_{\rm max}} \mathrm{d} \alpha \Delta_{\alpha}^{\rm T},
%\label{eq:S_FL}
%\end{eqnarray}
where $\Delta_{\alpha}^{\rm T}$ is defined by
\begin{equation}
\Delta_{\alpha}^{\rm T} = \frac{1}{N} \overline{\left\langle \sum_{i=1}^{N} | {\bf r}_i - {\bf r}_{{\rm ref}, i} |^2 \right\rangle_{\alpha}^{\rm T}}.
\end{equation}
To perform the integration and take the limit $\alpha_{\rm min} \to 0$ in Eq.~(\ref{eq:S_FL_def}), we write:
\begin{equation}
\lim_{\alpha_{\rm min} \to 0} \int_{\alpha_{\rm min}}^{\alpha_{\rm max}} \mathrm{d} \alpha \Delta_{\alpha}^{\rm T}  \simeq \alpha_{\rm min}  \Delta_{\alpha_{\rm min}}^{\rm T}  + \int_{\alpha_{\rm min}}^{\alpha_{\rm max}} \mathrm{d} \alpha \Delta_{\alpha}^{\rm T}.
\label{eq:integration_Delta}
\end{equation}
The practical choice for $\alpha_{\rm max}$ is simple, as it is sufficient that it lies deep inside the Einstein solid regime where $\Delta_{\alpha}^{\rm T} = d/(2\alpha)$ is satisfied. For $\alpha_{\rm min}$, a more careful look at the simulation results is needed. 

In Fig.~\ref{fig:KA}(c), we show $\Delta_{\alpha}^{\rm T}$ for the Kob-Andersen model at low temperature. The Einstein solid limit is satisfied for large $\alpha$, and we can fix $\alpha_{\rm max}= 10^7$. When $\alpha$ decreases, deviations from Einstein solid behavior are observed, and a plateau emerges. 
%With decreasing $\alpha$, $\Delta_{\alpha}^{\rm T}$ enters a plateau region shown by the shaded region. In this region, the system is trapped by its own cage.
Decreasing $\alpha$ further, the harmonic constraint for $\Delta_{\alpha}^{\rm T}$ becomes too weak and the glass metastability is not sufficient to prevent the system from diffusing away from the reference configuration, which translates into an upturn of $\Delta_{\alpha}^{\rm T}$ at small $\alpha$. 
 %of the basin is not strong enough to prevent the system from diffusing, which translates into an upturn of $\Delta_{\alpha}^{\rm T}$ for higher temperature at small $\alpha$.
It is instructive to compare the plateau level with the Debye-Waller factor measured from the bulk dynamics,~\cite{coslovich2018dynamic} indicated by a dashed line. 
This comparison shows that $\alpha_{\rm min} \approx 2$ is a good compromise: it is in the middle of the plateau, and corresponds to vibrations comparable to the ones observed in the bulk.
%The horizontal dashed line is the plateau height of independently measured (dynamic) mean-squared displacement, $\Delta(t)=\frac{1}{N} \langle \sum_{i=1}^N | {\bf r}_i(t) - {\bf r}_i(0) |^2 \rangle_{\alpha=0}^{\rm T}$, of the target system at the lowest temperature~\cite{coslovich2018dynamic}.
%The plateau height of $\Delta(t)$ matches well with $\Delta_{\alpha}^{\rm T}$ around $\alpha_{\rm min}$, suggesting that the Frenkel-Ladd method precisely takes into account the entropy of the vibrational motion inside a basin~\cite{ozawa2018ideal}.
Using this value for $\alpha_{\rm min}$, we obtain the Frenkel-Ladd glass entropy shown in Fig.~\ref{fig:KA}(a). We observe that $S_{\rm glass}^{\rm FL}$ is smaller than $S_{\rm glass}^{\rm harm}$, and becomes comparable to the anharmonic estimate using inherent structures, $S_{\rm glass}^{\rm anh}$, as temperature decreases, confirming that anharmonicities are automatically captured by the Frenkel-Ladd method.~\cite{ozawa2018ideal}

We show the resulting $S_{\rm conf}=S_{\rm tot}-S_{\rm glass}^{\rm FL}$ in Fig.~\ref{fig:kauzmann}.
Comparing with experimental data, the temperature range where $S_{\rm conf}$ can be measured is limited since the SWAP algorithm is not efficient for binary mixtures such as the Kob-Andersen model.~\cite{flenner2006hybrid}
Nevertheless an extrapolation to lower temperature suggests that $S_{\rm conf}/N$ may vanish at a finite $T_{\rm K}$.~\cite{ozawa2018configurational}

\subsection{Mixing entropy in the glass state}

\label{sec:mixing}

\begin{figure}
\includegraphics[width=0.95\columnwidth]{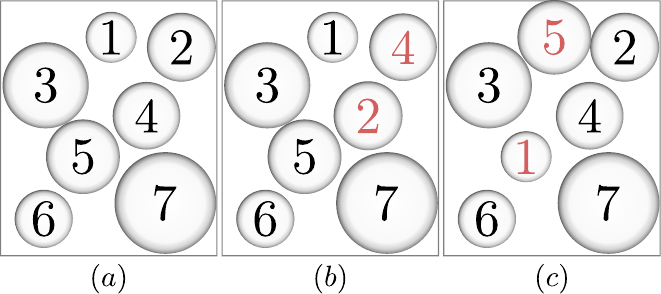}
\caption{Mixing entropy conundrum for continuous polydispersity. Should one treat these 3 configurations as 3 distinct glass states, or only two by grouping (a) and (b) together? In Sec.~\ref{sec:GFL}, a computational measurement is described that provides the correct answer, instead of guessing it.}
\label{fig:s_mix}
\end{figure}  

%@@ mention the major conceptual difference between a global constraint (on the density) versus a local constraint (on single particle displacements) @@

Using multi-component mixtures is essential to study supercooled liquids and glasses for spherical particle systems, as exemplified by metallic~\cite{chen2011brief} and colloidal~\cite{hunter2012physics,gokhale2016deconstructing} glasses.
This is also true for most computer simulations, since monocomponent systems crystallize too easily, except for large spatial dimensions~\cite{CIPZ11} or exotic mean-field like model systems.~\cite{IM11} For such multi-components systems, a mixing entropy term appears in the total entropy, see Eq.~(\ref{eq:entropy_ideal_gas}), with no analog in the glass entropy, see Eqs.~(\ref{eq:S_harm}) and (\ref{eq:S_FL}). Physically, this is because we decided to treat two configurations where distinct particles had been exchanged as two distinct glass states. 

%In the multi-component mixture, the mixing entropy has to be properly taken into account. The total entropy contains the mixing entropy contribution from Eq.~(\ref{eq:entropy_ideal_gas}), whereas we did not take into account it in the entropy of glass state shown in Eqs.~(\ref{eq:S_harm}) and (\ref{eq:S_FL}).

%This treatment originates from the following picture: Particles with different species can be exchanged their positions in the liquid state, whereas this process is not allowed in the glass state, because each particle is trapped by the cage. Consequently, the configurational entropy of the system contains the mixing entropy contribution inherited from the total entropy in the standard computational method~\cite{SKT99,sastry2000evaluation,sastry2001relationship,flenner2006hybrid,banerjee2014role,ozawa2015equilibrium}.

For typical binary mixtures studied in computer simulations, the mixing entropy is about as large as the configurational entropy itself over the range accessible to molecular dynamics simulations.~\cite{SKT99,AF07} Therefore, neglecting the mixing entropy can change the configurational entropy by about $100\%$, which in turn produces a similar uncertainty on the estimate of the Kauzmann temperature. Properly dealing with the mixing entropy is thus mandatory.~\cite{OB17}

%The order of magnitude of $S_{\rm mix}^{(M)}$ is comparable to $S_{\rm conf}$ at lower temperature~\cite{SKT99} and higher density~\cite{AF07}, thus neglecting the mixing entropy contributions causes huge underestimation of $S_{\rm conf}$, thus the estimated location of $T_{\rm K}$ will be significantly~\cite{OB17}. Thus we need to take into account the mixing entropy contributions appropriately in order to get a good estimation of $T_{\rm K}$. 

For discrete mixtures, such as binary and ternary mixtures, with large size asymmetries, the above treatment produces an accurate determination of $S_{\rm conf}$.~\cite{SKT99,sastry2000evaluation,sastry2001relationship,flenner2006hybrid,banerjee2014role,ozawa2015equilibrium} However, for systems with a continuous distribution of particle sizes, such as colloidal particles and several computer models, this leads to unphysical results. In the liquid, the mixing entropy is formally divergent, since for $M=N$ it becomes 
$S_{\rm mix}^{(M=N)}/N=(\ln N!)/N \simeq \ln N -1  \to \infty$.~\cite{salacuse1982polydisperse,sollich2001predicting} Because the glass entropy remains finite in conventional treatments, the configurational entropy also diverges, leading to the conclusion that no entropy crisis can take place in systems with continuous polydispersity.~\cite{OB17,baranau2017another} A similar argument was proposed by Donev {\it et al} to suggest that an entropy crisis does not exist in binary mixtures.~\cite{DTS06}

In fact, the above treatments do not accurately quantify the mixing entropy contribution in the glass entropy. This can be easily seen by considering a continuously polydisperse material with a very narrow size distribution, which should physically behave as a mono-component system, but has a mathematically divergent mixing entropy. In addition to this trivial example, the fundamental problem is illustrated in Fig.~\ref{fig:s_mix}, which sketches three configurations which differ by the exchange of a single pair of particles. The inherent structure and the standard Frenkel-Ladd methods treat those three configurations as distinct. Physically, configurations (a) and (b) should instead be considered as the same glass state, since they differ by the exchange of two particles with nearly identical diameters. The glass entropy should contain some amount of mixing entropy, taking into account those particle permutations that leave the glass state unaffected.~\cite{OB17} 

Recently, two methods were proposed to estimate the glass mixing entropy. The first method provides a simple approximation to the glass mixing entropy using information about the potential energy landscape.~\cite{OB17} We describe the second one in the next subsection, which leads to a direct determination of the glass mixing entropy using a generalized Frenkel-Ladd approach.~\cite{ozawa2018configurational} 

\subsection{Generalized Frenkel-Ladd method to measure the glass mixing entropy}

\label{sec:GFL}

A proper resolution to the problematic glass mixing entropy is to directly measure the amount of particle permutations allowed by thermal fluctuations, instead of making an arbitrary decision.~\cite{ozawa2018configurational}   
%
%The above physical resolusion of the problem of mixing entropy can be mathematically demonstrated by a generalization of the standard Frenkel-Ladd method~\cite{ozawa2018configurational}.
Technically, one needs to include particle permutations in the statistical mechanics treatment of the system. 
%The key point is including the permuation of the diameters as an additional degrees of freedom, which enables us to obtain precise estimation of the glass entropy including relevant mixing entropy contributions.
In addition to the positions, we introduce the particle diameters, represented as $\Sigma^N=\{ \sigma_1, \sigma_2, \cdots, \sigma_N  \}$.
Let $\pi$ denote a permutation of $\Sigma^N$, and
$\Sigma_{\pi}^N$ represents the resulting sequence. There exist $N!$ such permutations. We define a reference sequence $\Sigma_{\pi^*}^N=(\sigma_1, \sigma_2, \sigma_3, \cdots, \sigma_N)$. The potential energy now depends on both positions and diameters, $U (\Sigma_{\pi}^N, {\bf r}^N)$. For simplicity, we write $U ({\bf r}^N)=U (\Sigma_{\pi^*}^N, {\bf r}^N)$ for the reference $\Sigma_{\pi^*}^N$.
% and drop off $\Sigma_{\pi^*}^N$ from the argument.

Including particle diameters as additional degrees of freedoms, the partition function reads
\begin{equation}
\mathcal{Z}=  \frac{1}{N!} \sum_{\pi} \frac{\Lambda^{-Nd}}{\Pi_{m=1}^M N_m! } \int_V \mathrm{d} {\bf r}^N  e^{-\beta U (\Sigma_{\pi}^N, {\bf r}^N)}.
\label{eq:Z_new}
\end{equation}
This partition function is the correct starting point to compute the configurational entropy in multi-component systems, including continuous polydispersity. The resulting method is a straightforward generalization of the Frenkel-Ladd method.~\cite{frenkel1984new}

As before, we introduce a reference configuration and a harmonic constraint, 
\begin{equation}
\beta U_{\alpha}(\Sigma_{\pi}^N, {\bf r}^N, {\bf r}_{\rm ref}^N) = \beta U(\Sigma_{\pi}^N, {\bf r}^N) + \alpha \sum_{i=1}^{N} | {\bf r}_i - {\bf r}_{{\rm ref}, i} |^2,
\label{eq:hamiltonian}
\end{equation}
where ${\bf r}_{\rm ref}^N$ is a reference equilibrium configuration.

For the unconstrained system with $\alpha=0$, the partition function in Eq.~(\ref{eq:Z_new}) reduces to the conventional partition function in Eq.~(\ref{eq:Z_old}), because diameter permutations can be compensated by the configurational integral. Therefore, the computation of $S_{\rm tot}$ is not altered by the introduction of the permutations. For the glass state with $\alpha > 0$, the partition function in Eq.~(\ref{eq:Z_new}) and the corresponding statistical average become
\begin{eqnarray}
\mathcal{Z}_{\alpha} &=&  \frac{1}{N!} \sum_{\pi} \frac{N!\Lambda^{-Nd}}{\Pi_{m=1}^M N_m! } \int_V \mathrm{d} {\bf r}^N e^{-\beta U_{\alpha}(\Sigma_{\pi}^N, {\bf r}^N, {\bf r}_{\rm ref}^N)}. \label{eq:Z_alpha} \\
\left\langle (\cdots) \right\rangle_{\alpha}^{\rm T, S} &=& \frac{ \sum_{\pi} \int_V \mathrm{d} {\bf r}^N  (\cdots) e^{ -\beta U_{\alpha}(\Sigma_{\pi}^N, {\bf r}^N, {\bf r}_{\rm ref}^N)}  }{ \sum_{\pi} \int_V \mathrm{d} {\bf r}^N  e^{-\beta U_{\alpha}(\Sigma_{\pi}^N, {\bf r}^N, {\bf r}_{\rm ref}^N)}}, \label{eq:T_S}
\end{eqnarray}
We add a factor $N!$ in the numerator of Eq.~(\ref{eq:Z_alpha}), because 
there exist $N!$ configurations defined by the permutations of the particle identities of the reference configuration ${\bf r}_{\rm ref}^N$. More crucially, due to the presence of ${\bf r}_{\rm ref}^N$, the partition function in Eq.~(\ref{eq:Z_alpha}) is not identical to the one in Eq.~(\ref{eq:zalpha}). 

Following the same steps as before we get the glass entropy by a generalized Frenkel-Ladd method, defined as
\begin{eqnarray}
S_{\rm glass}^{\rm GFL} &=&  S_{\rm \alpha_{\rm max}}  + N \lim_{\alpha_{\rm min} \to 0} \int_{\alpha_{\rm min}}^{\alpha_{\rm max}} \mathrm{d} \alpha \Delta_{\alpha}^{\rm T,S} \nonumber \\
&\quad& + S_{\rm mix}^{(M)} -  \overline{ \mathcal{S}_{\rm mix}({\bf r}_{\rm ref}^N, \beta)},
\label{eq:S_glass_final}
\end{eqnarray}
with 
\begin{equation}
\Delta_{\alpha}^{\rm T, S} = \frac{1}{N} \overline{\left\langle \sum_{i=1}^{N} | {\bf r}_i - {\bf r}_{{\rm ref}, i} |^2 \right\rangle_{\alpha}^{\rm T,S}},
\end{equation}
and $\mathcal{S}_{\rm mix}({\bf r}_{\rm ref}^N, \beta)$ is obtained as
\begin{equation}
\mathcal{S}_{\rm mix}({\bf r}_{\rm ref}^N, \beta) =  - \ln \left( \frac{1}{N!} \sum_{\pi} e^{-\beta \left( U(\Sigma_{\pi}^N, {\bf r}_{\rm ref}^N) - U({\bf r}_{\rm ref}^N) \right) } \right).
\label{eq:s_mix_integral}
\end{equation}
%where $\Delta U_{\rm mix}({\bf r}_{\rm ref}^N, \beta')$ is the energy increment associated with diameter exchanges at fixed positions ${\bf r}_{\rm ref}^N$. 
%\begin{eqnarray}
%\Delta U_{\rm mix}({\bf r}_0^N, \beta') &=&  \left\langle U_0(\Sigma_{\pi}^N, {\bf r}_0^N) \right\rangle_{\beta'}^{\rm S} - U_0({\bf r}_0^N), \\
%\left\langle (\cdots) \right\rangle_{\beta}^{\rm S} &=& \frac{ \sum_{\pi} (\cdots) e^{ -\beta U_0(\Sigma_{\pi}^N, {\bf r}_0^N) }  }{  \sum_{\pi}  e^{ -\beta U_0(\Sigma_{\pi}^N, {\bf r}_0^N) }}.
%\label{eq:S}
%\end{eqnarray}
Note that in Eq.~(\ref{eq:S_glass_final}), the mean-squared displacement $\Delta_{\alpha}^{\rm T,S}$ is evaluated by simulations where both positions and diameters fluctuate, and we expect $\Delta_{\alpha}^{\rm T,S} \geq \Delta_{\alpha}^{\rm T}$. Practically, $\Delta_{\alpha}^{\rm T,S}$ is computed by Monte Carlo simulations including standard translational displacements and diameter swaps.
%In addition, since particle permutations allowed by thermal fluctuations are now performed, we expect that particle exchanges of particles with similar diameters arise more frequently than for asymmetric sizes, so that $S_{\rm glass}^{\rm GFL}$ correctly measures the mixing entropy contribution to the glass entropy. 
%Figure~\ref{fig:GFL}(a) for the polydisperse soft spheres shows $\Delta_{\alpha}^{\rm T,S} > \Delta_{\alpha}^{\rm T}$ around the plateau region, suggesting that $\Delta_{\alpha}^{\rm T,S}$ takes into account a wider range of the phase space volume than purely vibrational contribution, $\Delta_{\alpha}^{\rm T}$~\cite{ozawa2018ideal}. Hence it is expected that the resulting $S_{\rm GFL}$ more correctly deals with the non-vibrational contributions to the glass entropy.
In addition to this, $S_{\rm glass}^{\rm GFL}$ in Eq.~(\ref{eq:S_glass_final}) contains another non-trivial contribution,
$S_{\rm mix}^{(M)} - \overline{ \mathcal{S}_{\rm mix}}$, which requires Monte Carlo simulations of the diameter swaps for a fixed ${\bf r}_{\rm ref}^N$. In practice the entropy in Eq.~(\ref{eq:s_mix_integral}) is evaluated by a thermodynamic integration.~\cite{ozawa2018configurational}

%To measure $\Delta U_{\rm mix}({\bf r}_0^N, \beta')$ in practice, the system is gradually heated from the target temperature $\beta'=\beta$ to the infinite temperature $\beta' \to 0$ by performing Monte Carlo simulations where only particle diameter permutations (and hence products of diameter swaps) are attempted while keeping fixed the particle positions of the reference configuration ${\bf r}_0^N$ generated at $\beta$.

For mixtures with large size asymmetry such as the Kob-Andersen model, particle permutations of un-like particles rarely happen,~\cite{flenner2006hybrid} and the generalized Frenkel-Ladd method yields 
%For monodisperse particles or discrete mixtures where the swap of the diameters with different species have a high energy cost~\cite{flenner2006hybrid}, the equalities,  
$\overline{ \mathcal{S}_{\rm mix}}=S_{\rm mix}^{(M)}$ and $\Delta_{\alpha}^{\rm T,S} = \Delta_{\alpha}^{\rm T}$,  
%, would hold, as numerically confirmed in Fig.~\ref{fig:GFL}(b) for the Kob-Andersen model.
so that Eq.~(\ref{eq:S_glass_final}) reduces to the conventional Frenkel-Ladd method in Eq.~(\ref{eq:S_FL}). 
On the other hand, for continuously polydisperse systems or mixtures with small size asymmetry, we expect $\overline{ \mathcal{S}_{\rm mix}}/N < S_{\rm mix}^{(M)}/N \to \infty$ and $\Delta_{\alpha}^{\rm T,S} > \Delta_{\alpha}^{\rm T}$. 
In the limit case of a very narrow continuous distribution, we would have 
$\overline{ \mathcal{S}_{\rm mix}}/N = 0$ and $\Delta_{\alpha}^{\rm T,S} = \Delta_{\alpha}^{\rm T}$, and we automatically get back to the treatment of a mono-component material. 

%Indeed, $\overline{ \mathcal{S}_{\rm mix}}$ in Fig.~\ref{fig:GFL}(b) takes finite values for systems with continuous polydispersity.

%Therefore Eq.~(\ref{eq:S_glass_final}) is a straightforward generalization of the conventional Frenkel-Ladd method for systems with continuous polydispersity, and the thermodynamic integration automatically takes into account the correct number of permutations allowed by thermal fluctuations in equilibrium~\cite{ozawa2018configurational}.  

We finally obtain the configurational entropy as $S_{\rm conf}=S_{\rm tot}-S_{\rm glass}^{\rm GFL}$, which finally resolves the paradox raised by the mixing entropy in conventional schemes. For polydisperse systems, both the total entropy and the glass entropy in Eq.~(\ref{eq:S_glass_final}) contain the diverging mixing entropy term, which thus cancel each from the final expression of the configurational entropy. Instead, the physical mixing entropy contribution is quantified by $ \overline{\mathcal{S}_{\rm mix}({\bf r}_{\rm ref}^N, \beta)}$ in Eq.~(\ref{eq:s_mix_integral}), which is finite, and whose value depends on the detailed particle size distribution of the system. 
 
%It should be obvious that the present scheme resolves the problem of an infinite mixing entropy for continuous polydispersity~\cite{OB17}. The diverging mixing entropy is the term $S_{\rm mix}^{(M)}$ in $S_{\rm tot}$ (through Eq.~(\ref{eq:entropy_ideal_gas})) which appears also in $S_{\rm GFL}$ in Eq.~(\ref{eq:S_glass_final}). Instead $\overline{\mathcal{S}_{\rm mix}}$ in Eq.~(\ref{eq:S_glass_final}) remains as a finite mixing entropy contribution to $S_{\rm conf}$. Thus, the configurational entropy automatically incorporates the correct information about size polydispersity. Besides, the thermodynamic integration of $\Delta_{\alpha}^{\rm T,S}$ enables the system explore wider range of the phase space volume than purely vibrational description of the glass state, suggesting that the nature of the glass state composing of many inherent structures is properly taken into account in this scheme.

\begin{figure}
\includegraphics[width=0.95\columnwidth]{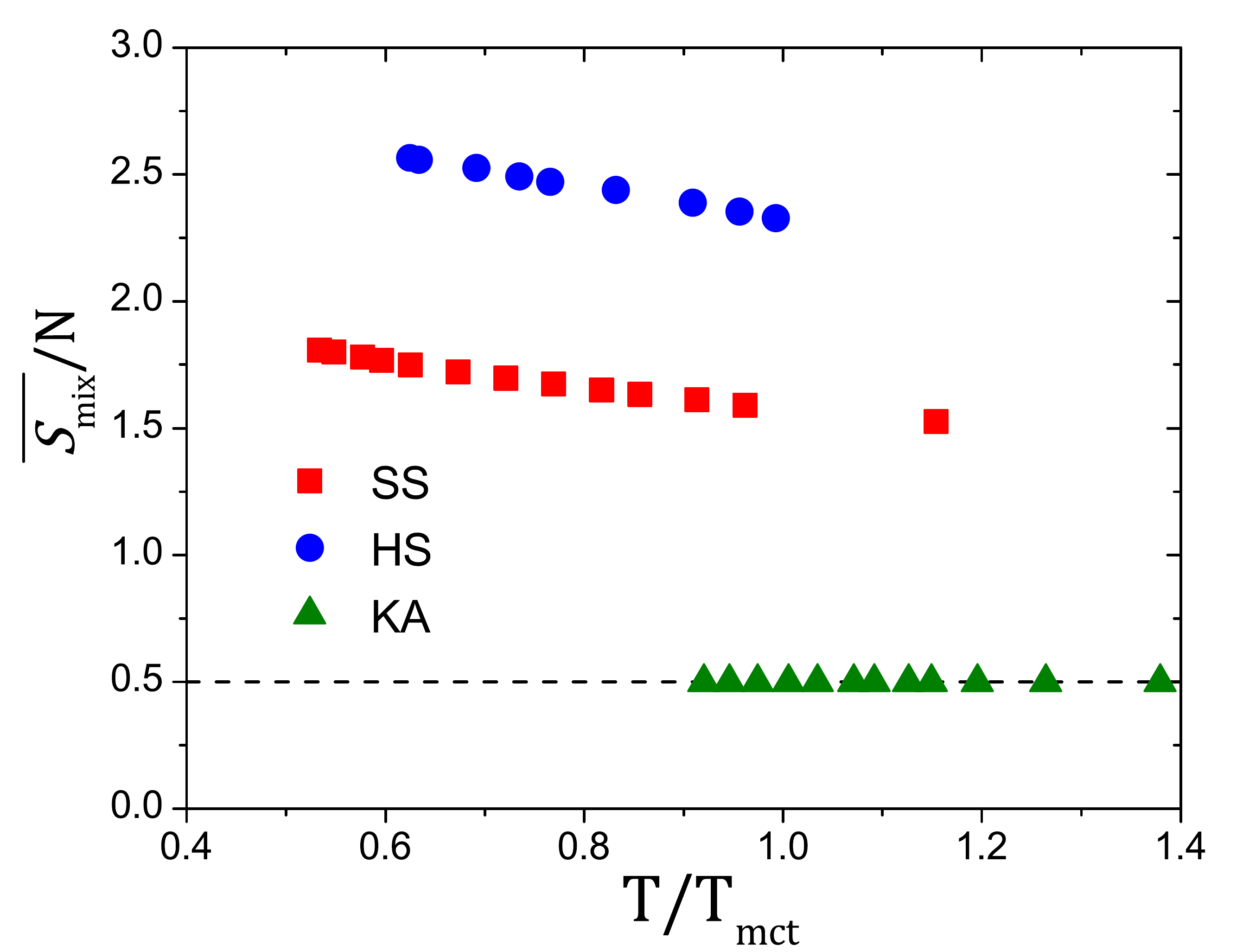}
\caption{Mixing entropy $\overline{\mathcal{S}_{\rm mix}}/N$ as a function of the normalized temperature $T/T_{\rm mct}$ for polydisperse soft spheres (SS), hard spheres (HS), and the Kob-Andersen model (KA).
The dashed-line corresponds to the combinatorial mixing entropy for the KA mixture.}
\label{fig:GFL}
\end{figure}  

In Fig.~\ref{fig:GFL}, we show the measured $ \overline{\mathcal{S}_{\rm mix}({\bf r}_{\rm ref}^N, \beta)}$ for three representative glass-forming models. For the Kob-Andersen binary mixture, the combinatorial mixing entropy, $S_{\rm mix}^{(M=2)}/N \approx 0.5$,~\cite{SKT99,sastry2000evaluation} is found, whereas for continuously polydisperse soft~\cite{NBC17} and hard spheres~\cite{berthier2016equilibrium} with polydispersity $\approx 23\%$, a finite value of the mixing entropy is measured, with a non-trivial temperature dependence. The data also directly confirms that the mixing entropy cannot be used to disprove the existence of a Kauzmann transition.~\cite{DTS06} 
 
Figure~\ref{fig:kauzmann} shows the final result, $S_{\rm conf}=S_{\rm tot}-S_{\rm glass}^{\rm GFL}$, for polydisperse hard and soft spheres along isochoric~\cite{ozawa2018configurational} and isobaric paths (in preparation), in $d=2$~\cite{berthier2018zero} and $3$.
For the hard sphere model, we use the inverse of the reduced pressure, $1/p=\rho T/P$, as the analog of the temperature.
Thanks to the \res{efficiency} of the SWAP algorithm for these models, we can measure a reduction of the configurational entropy comparable to experimental molecular liquids, and even access values measured in vapor deposited ultrastable glasses.~\cite{fullerton2017density} 
Therefore, the simulation results presented here, together with experimental ones, offer the most complete and persuasive data set for existence of the Kauzmann transition at a finite temperature in $d=3$ and at zero temperature in $d=2$.
 
\section{Configurational entropy from free energy landscape}

\label{sec:FP}

%The results based on the generalized Frenkel-Ladd schem intrigues us to consider amorphous ``states'' as a minimum of the free energy landscape as a function of the denisty profile, which would contain a bunch of inherent structures~\cite{biroli2000inherent}. In order to handle these exponential number of glass states at once, it is convinient to introduce the order paramer defining liquid and glass states, and discuss its free energy, akin to the standard Landau free energy approach.

\subsection{Franz-Parisi Landau free energy}

The mean-field theory of the glass transition introduced in Sec.~\ref{sec:MF} suggests a well-defined route to the configurational entropy,~\cite{BC14} based on free energy measurements of a Landau free energy $V(Q)$, expressed as a function of the overlap $Q$ between pairs of randomly chosen equilibrium configurations.~\cite{FP95,FP97} A practical definition of the overlap was given in Eq.~(\ref{eq:overlap}). The introduction of the appropriate global order parameter to detect the glass transition driven by an entropy crisis is the first key step.

The second key point is the assumption that $V(Q)$ contains, for finite dimensional systems, the relevant information about the configurational entropy. As illustrated in Fig.~\ref{fig:FPMF}, mean-field theory suggests that the glass phase at large $Q$, for $T_K < T < T_{\rm mct}$, is metastable with respect to the equilibrium liquid phase at small $Q$, with a free-energy difference between the two phases that is controlled by the configurational entropy. To measure this configurational entropy, one should first demonstrate the existence of the glass metastability, and use it to infer $S_{\rm conf}$ as a free energy difference between liquid and glass phases.

The computational tools to study $V(Q)$ and metastability are not specific to the glass problem, but can be drawn from computer studies of ordinary first-order phase transitions.~\cite{frenkel2001understanding} To analyze the overlap and its fluctuations, we introduce a reference equilibrium configuration ${\bf r}_{\rm ref}^N$. We then define the overlap $Q_{\rm ref}=Q({\bf r}^N,{\bf r}_{\rm ref}^N)$ between the studied system ${\bf r}^N$ and the reference configuration, and introduce a field, $\varepsilon$, conjugate to the overlap, 
\begin{equation}
U_{\varepsilon}({\bf r}^N,{\bf r}_{\rm ref}^N)=U({\bf r}^N) - \varepsilon N Q({\bf r}^N,{\bf r}_{\rm ref}^N),
\label{eq:Hamiltonian_FP}
\end{equation}
where $U$ is the potential energy of the unconstrained bulk system ($\varepsilon = 0$). The corresponding statistical mechanics and average become 
\begin{eqnarray}
Z_{\varepsilon} &=&  \Lambda^{-Nd} \int_V \mathrm{d} {\bf r}^N  e^{-\beta U_{\varepsilon}({\bf r}^N, {\bf r}_{\rm ref}^N)}, \\
\left\langle (\cdots) \right\rangle_{\varepsilon} &=& \frac{\int_V \mathrm{d} {\bf r}^N  (\cdots) e^{ - \beta U_{\varepsilon}({\bf r}^N,{\bf r}_{\rm ref}^N)}}{\int_V \mathrm{d} {\bf r}^N e^{ - \beta U_{\varepsilon}({\bf r}^N, {\bf r}_{\rm ref}^N)}}, \label{eq:ave_epsilon} 
\label{eq:partfuncepsilon}
\end{eqnarray}
%Note that we do not need to include the factorial term, because the Franz-Parisi potential approach directly associates the free energy ``difference'' between liquid and glass states, hence the factorial term and thus the comibinatrial mixing entroy $S_{\rm mix}^{(M)}$ cancells out automatically.
and the related Helmholtz free energy is obtained as
\begin{equation}
-\beta F_{\varepsilon} = \overline{\ln Z_{\varepsilon}},
\end{equation}
where the overline denotes an average over independent reference configurations. All thermodynamic quantities can then be deduced from $F_{\varepsilon}$, such as the average overlap $\langle Q \rangle_\varepsilon = - (1/N) \partial F_\varepsilon / \partial \varepsilon$.

%The averaged overlap $\overline{\left\langle Q_{12} \right\rangle_{\varepsilon}}$ behaves essentially similar to $m$ vs. $h$ curves, showing a liquid to glass transition driven by the external field $\varepsilon$. We show $\overline{\left\langle Q_{12} \right\rangle_{\varepsilon}}$ vs. $\varepsilon$ for the polydisperse hard spheres in Fig~\ref{fig:FP}(a). For lower density (higher temperature), $\overline{\left\langle Q_{12} \right\rangle_{\varepsilon}}$ monotonically increases steadly with $\varepsilon$ from liquid state ($\overline{\left\langle Q_{12} \right\rangle_{\varepsilon}} \approx 0.05$) to glass state ($\overline{\left\langle Q_{12} \right\rangle_{\varepsilon}} \approx 0.8$). For suffiently high density (low temperature), $\overline{\left\langle Q_{12} \right\rangle_{\varepsilon}}$ discontinuously jump as a first-order transition from liquid to glass states at the critical conjugate field $\varepsilon^{\star}$. Therefore it is evident that $Q_{12}$ has a standard character of phase transition.

Following the spirit of the Landau free energy,~\cite{chaikin1995principles} we express the free energy as a function of the order parameter $Q$, instead of $\varepsilon$. The  Franz-Parisi free energy $V(Q)$ is obtained by a Legendre transform of $F_\varepsilon$, 
%Following the standard Laudau free energy approach, it is convinient to use the free energy as a funciton of the order parameter in order to discuss the phase transition in the bulk (without field). To this end, we define the Franz-Parisi potential $V(Q)$ by the Legendre transform of $F_{\varepsilon}$ as %Gibbs free energy $G(m)$ or Landau free energy is obtained by the Legendre transform of the Helmholtz free energy $F(h)$ as $G(m)/N=\min_{h} \{ F(h)/N + m h\}$
\begin{equation}
V(Q) =\frac{1}{N} \left( \min_{\varepsilon}\left\{ F_{\varepsilon} + \varepsilon N Q \right\} -F_0 \right),
\label{eq:VQ_def}
\end{equation}
where $F_0=-\beta^{-1}\ln Z_0$ is the free energy of the unconstrained system, which simply ensures that $V(Q) = 0$ for the equilibrium liquid phase at small $Q$. Following standard computational approaches for free-energy calculations,~\cite{frenkel2001understanding} $V(Q)$ is directly obtained by probing the equilibrium fluctuations of the overlap, 
\begin{eqnarray}
V(Q)&=&-\frac{T}{N} \overline{\ln \frac{ \Lambda^{-Nd}}{Z_0 } \int_V \mathrm{d} {\bf r}^N  e^{-\beta U({\bf r}^N)} \delta(Q-Q_{\rm ref})}, \nonumber \\
&=& -\frac{T}{N} \overline{\ln P(Q)},
\label{eq:defF}
\end{eqnarray}
where $P(Q)=\langle \delta(Q-Q_{\rm ref}) \rangle$ is the probability distribution of the overlap function for the unconstrained bulk system.

This method naturally solves issues about the mixing entropy.~\cite{OB17} As captured in Eq.~(\ref{eq:VQ_def}), this construction treats free energy differences, with no need to define absolute values for the entropy. The combinatorial terms in Eq.~(\ref{eq:partfuncepsilon}) are therefore not included since they eventually cancel out.
Additionally, the constraint applied to the system acts only on the value of the overlap $Q$. Since particle permutations do not affect the value of the overlap, see Eq.~(\ref{eq:overlap}), particle permutations \res{within the same species} can occur both in the liquid, near $Q_{\rm liq}$, and in the glass, near $Q_{\rm glass}$, with a probablity controlled by thermal fluctuations. 

In finite dimensions, the secondary minimum in $V(Q)$ obtained in the mean-field limit (see Fig.~\ref{fig:FPMF}) cannot exist, as the free energy must be convex, for stability reasons.~\cite{Call85} At best, $V(Q)$ should develop a small non-convexity for finite system sizes, and a linear part for larger systems, as for any first-order phase transition. In the presence of a finite field $\varepsilon$, a genuine first-order liquid-to-glass transition is predicted,~\cite{FP95,FP97} where $\langle Q \rangle_{\varepsilon}$ jumps discontinuously to a large value as $\varepsilon$ is increased. This phase transition exists in the mean-field limit, and can in principle survive finite dimensional fluctuations.

The existence of this constrained phase transition induced by a field $\varepsilon$ in finite dimensional systems is needed to identify the Franz-Parisi free-energy with the configurational entropy in the unconstrained bulk system. If a metastable glass phase is detected in some temperature regime $T>T_K$, then it is possible to measure the free-energy difference between the equilibrium liquid and the metastable glass, namely $V(Q_{\rm glass})$. This quantity represents the entropic cost of localizing the system in a single metastable state: this is indeed the configurational entropy.    

\subsection{Computational measurement}

The free-energy $V(Q)$ in Eq.~(\ref{eq:defF}) is the central physical quantity to measure in computer simulations.~\cite{berthier2013,BC14,ceiling17} It follows from the measurement of rare fluctuations of the overlap, since $P(Q) \sim e^{-\beta N V(Q)}$. Measuring such rare fluctuations (indeed, exponentially small in the system size) in equilibrium systems is a well-known problem that has received considerable attention and powerful solutions in the context of equilibrium phase transitions,~\cite{frenkel2001understanding} such as umbrella sampling.
Physically, the idea is to perform simulations in an auxiliary statistical ensemble where the Boltzmann weight is biased by a known amount, and from which the unbiased canonical distribution is reconstructed afterwards.~\cite{ferrenberg1988new,BJ15} Combining this technique to the swap Monte Carlo~\cite{NBC17} and parallel tempering methods~\cite{hukushima1996exchange} to sample more efficiently the relevant fluctuations makes possible the numerical measurement of $V(Q)$ over a broad range of physical conditions. 

%For Eq.~(\ref{eq:def_s_conf_V}),  the main computational task is getting $V(Q)$ and hence $P(Q)$. Typically $P(Q)$ behaves as $e^{-\beta N V(Q)}$. Thus $P(Q)$ is exponentially small in $N$ at large $Q$. 
%This is really difficult computational task especially at higer $Q$.
%We bypass this difficulity by using some extended ensemble techniques, employing biasing by a known amount to get unlikely fluctuations, then unbiasing
%to estimate the true probability~\cite{frenkel2001understanding}.
%We also combine these techniques with swap MC~\cite{NBC17} and the parallel temperating algorithm~\cite{hukushima1996exchange}.
%To recover the bulk information, we use histogram reweighting methods~\cite{ferrenberg1988new} to recontruct the 
%free energy profile from multiple biased simulations all probing the same quantity. 
%Other more advanced methods such as multi-cannonical etc. would help more.

The same numerical techniques can also be used to probe the existence and physical properties of the phase transition induced by a field $\varepsilon$.
The $\varepsilon$-transition has given rise to number of theoretical and computational analysis, which conclude that the transition is present in spatial dimensions $d>2$.~\cite{berthier2013,BC14,parisi2014liquid,BJ15,JG16,charlotteplaquette} The phase transition emerges for temperatures lower than a critical temperature $T^*$, which is the analog of $T_{\rm onset}$ defined in the mean-field theory. For $T< T^*$, a first-order phase transition appears at a finite value $\varepsilon^*(T)$ of the field, where the overlap jumps discontinuously to a value $Q_{\rm glass}(T)$.   

The existence of the transition allows the quantitative determination of  the configurational entropy, namely 
\begin{equation}
S_{\rm conf} = \frac{N}{T} V(Q=Q_{\rm glass}).
\label{eq:def_s_conf_V}
\end{equation}
%Indeed the Kauzmann transition where $S_{\rm conf}/N=0$ is perfectly caputured by $V(Q_{\rm glass})=0$.
A nearly equivalent determination can be obtained directly from the properties of the constrained phase transition, since $\varepsilon^*$ represents the field needed to tilt the Franz-Parisi free energy and make the local minimum at large $Q$ become the global one.~\cite{BC14,ceiling17}
%As we have seen in Fig~\ref{fig:FP}(a), the external field $\varepsilon$ can induce the liquid to glass transition.
%This transtiion is can be viewed as tilling of $V(Q)$ to induce the transition, as given by $V(Q_{\rm glass})- \varepsilon^{\star} Q_{\rm glass} = 0$. 
Taking into account the small but positive $Q_{\rm liq}>0$, we can estimate the configurational entropy as 
\begin{equation}
S_{\rm conf} \simeq  \frac{N}{T} \varepsilon^* (Q_{\rm glass}-Q_{\rm liq}).
\label{eq:def_s_conf_epsilon}
\end{equation}

%Two estimation of the configurational entropy in Eq.~(\ref{eq:def_s_conf_V}) and (\ref{eq:def_s_conf_epsilon}) do not require special considerations for metastable states, especially time scale issue. These are automatically taken into account in the construction. 

%Some technical issues would be defining the overlap in Eq.~(\ref{eq:def_overlap_FP}), especially, the value of $a$, and value of $Q_{\rm glass}$ (this is would be less important).

%Another difinition, Eq.~(\ref{eq:def_s_conf_epsilon}), requires $\varepsilon^{\star}$ that is determined by the $\overline{\left\langle Q_{12} \right\rangle_{\varepsilon}}$ vs. $\varepsilon$ plot. To get this plot, the extended ensemble techniques plus the parallel temperating algorithm are also useful.

\begin{figure}
\includegraphics[width=0.95\columnwidth]{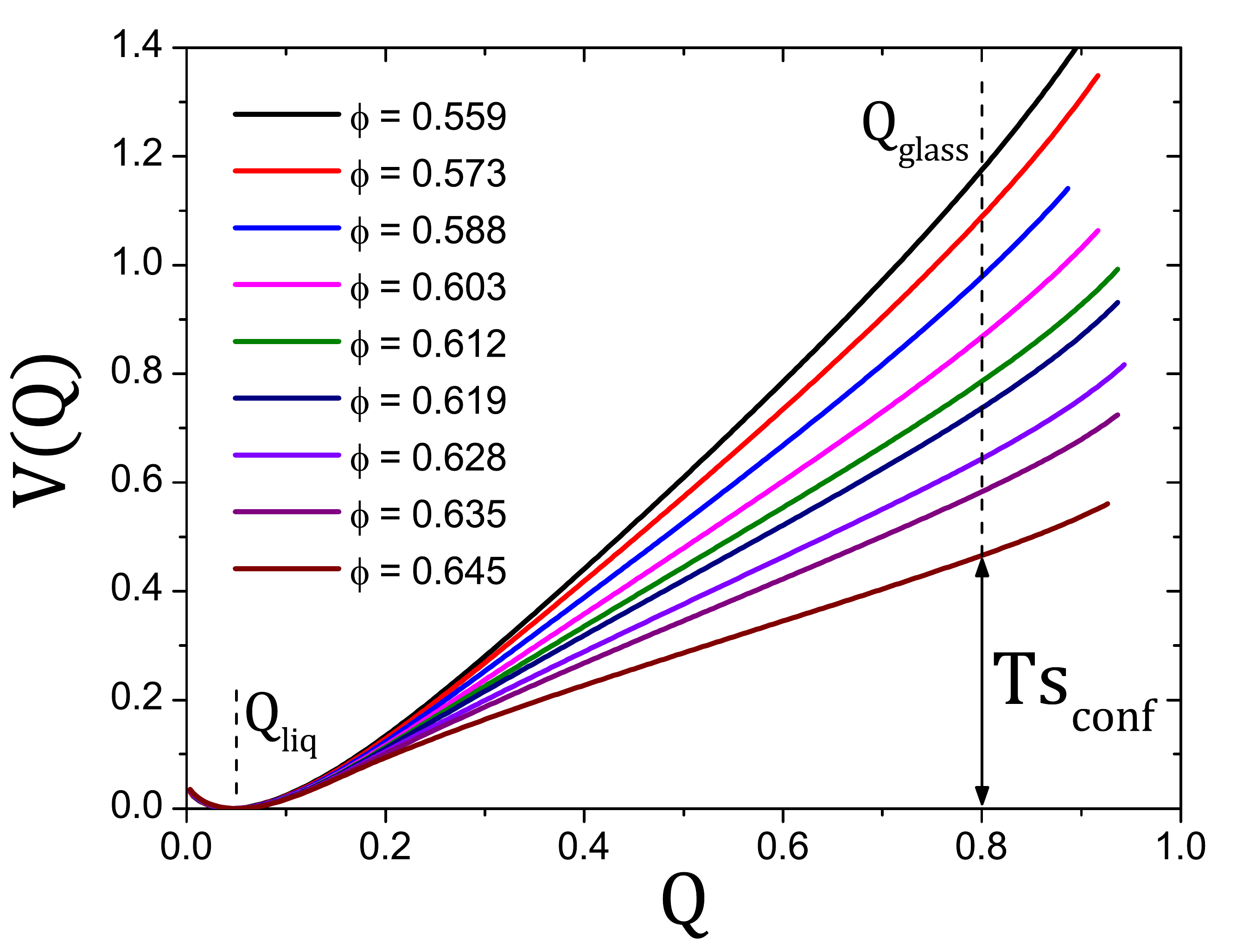}
\caption{Franz-Parisi free energy in 
three dimensional polydisperse hard spheres, using a combination of swap Monte Carlo, 
parallel tempering, and umbrella sampling techniques.~\cite{ceiling17} 
$V(Q_{\rm glass})$ decreases progressively with increasing the volume fraction $\phi$.
The vertical arrow indicates the estimate of $\SC$ as the free
energy difference between the low-overlap ($Q_{\rm liq}$) and large-overlap ($Q_{\rm glass}$) phases.
Estimated Kauzmann transition volume fraction, $\phi_{\rm K}$, at which $\SC$ vanishes is $\phi_{\rm K} \approx 0.68$.
The system shows jamming transition by rapid compression of dilute configurations at $\phi_{\rm J} \approx 0.655$.~\cite{ozawa2017exploring} 
}
\label{fig:FP}
\end{figure}  

In Fig.~\ref{fig:FP}, we show the evolution of the Franz-Parisi free energy $V(Q)$ for a system of continuously polydisperse hard spheres in three dimensions, with $N=300$ particles. The value of $Q_{\rm glass}$ is identified by a separate study of the $\varepsilon$-transition, and is indicated as a vertical dashed line. For each value of the volume fraction $\phi$, $V(Q_{\rm glass})$ provides an estimate of the configurational entropy using Eq.~(\ref{eq:def_s_conf_V}), as shown by the vertical arrow. 

More broadly, the data in Fig.~\ref{fig:FP} suggests that Kauzmann's intuition of an underlying thermodynamic phase transition connected to the rapid decrease of $S_{\rm conf}$ is realized in deeply supercooled liquid. \res{The evolution of the Franz-Parisi free energy shows that the glass phase at large $Q$ is metastable as $\phi < \phi_{\rm K}$ (i.e., $T>T_{\rm K}$), but its stability increases rapidly as $\phi$ increases (i.e., $T$ decreases), controlled by the decrease of the configurational entropy.} It is still not known whether a finite temperature entropy crisis truly takes place as a thermodynamic phase transition, but the key idea that glass formation is accompanied by the decrease of the associated free energy difference (and hence the configurational entropy) is no longer a hypothesis, but an established fact.  

Finally, the evolution of the free-energy $V(Q)$ with supercooling is quite dramatic. This large change quantitatively answers the question raised by the \res{apparent} similarity of the two particle configurations shown in Fig.~\ref{fig:cartoon}. The density profiles of those two state points \res{do not seem very different,} but their free energy profiles $V(Q)$ are. 
This means that to compare the two snapshots, one should monitor appropriate observables reflecting the reduction of available states in glass formation, instead of simple structural changes. 
 
\section{Configurational entropy from real space correlations}

\label{sec:PTS}

\subsection{A real space view of metastability}

In finite dimensions, the long-lived metastable states envisioned by mean-field theory do not exist since the system will eventually undergo structural relaxation in a finite time. Therefore, metastable states can at best exist over a finite timescale.~\cite{KTW89,biroli2001metastable} In the construction of Franz-Parisi,~\cite{FP95,FP97} metastable states are therefore explored by introducing a global constraint on the system via a field conjugate to the macroscopic overlap. This strategy allows one to estimate the number of free energy minima for a given temperature. 

The constraint envisioned in the Franz-Parisi is global and acts on the bulk system. In this section, we introduce another type of constraint that again allows a sharp distinction between the vicinity of a given configuration (the glass basin), and the rest of the free energy landscape. The key difference with the Franz-Parisi constraint is that we impose a spatially resolved constraint to the system using a cavity construction.~\cite{BB04} We shall argue that this provides a real space interpretation of the rarefaction of metastable states in terms of a growing spatial correlation length, the so-called point-to-set correlation length. This correlation length cannot emerge from the observation of the density profile in a single configuration, but stems once again from the comparison between the distinct density profiles available under some constraint. 

\begin{figure}
\includegraphics[width=0.75\columnwidth]{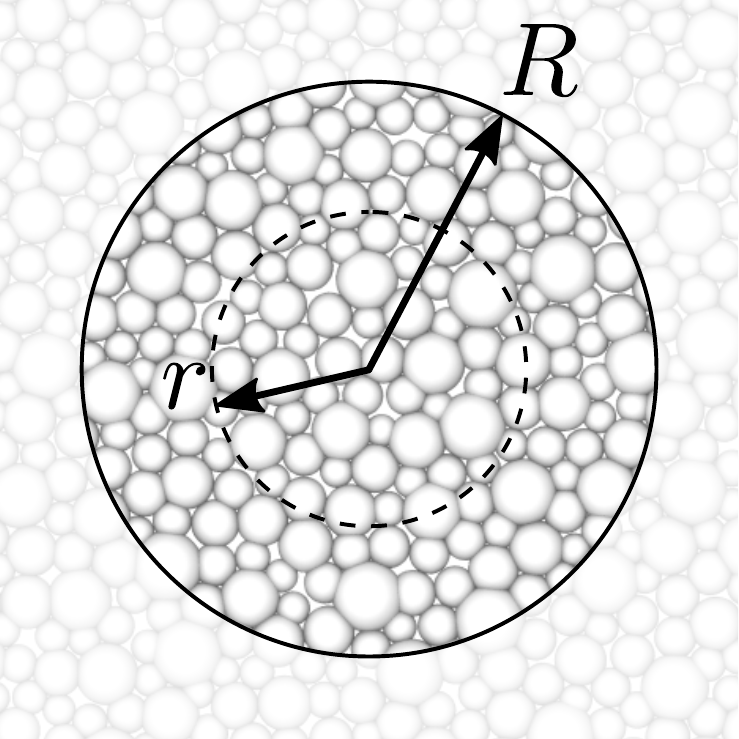}
\caption{Sketch of the cavity construction to determine the point-to-set lengthscale. The positions of the particles outside a cavity of radius $R$ are given by a reference equilibrium configuration and are frozen, while particles inside the cavity evolve freely at thermal equilibrium in the presence of the frozen amorphous boundaries. The overlap profile $Q(r)$ is defined by comparing the density profile inside the cavity to the reference configuration.}
\label{fig:pts_pic}
\end{figure}  

The main idea is illustrated in  Fig.~\ref{fig:pts_pic}. We prepare an equilibrium configuration of the system, which we take as a reference configuration, ${\bf r}_{\rm ref}^N$. We then consider a configuration ${\bf r}^N$ which is constrained to be equal to the reference configuration outside a cavity of radius $R$, but can freely fluctuate inside the cavity. Therefore, the constraint from the reference configuration is now only felt at the frozen amorphous boundary of the cavity. By varying the cavity size $R$, one can then infer how far the constraint propagates inside the cavity. As quantified below, one expects a crossover between small cavities where the constraint is so strong that particles inside the cavity can only remain close to the reference configuration, whereas for very large cavities particles inside the cavity will explore a large number of distinct states. The crossover between these two regimes is used to define the point-to-set correlation length.~\cite{BB04,pts1,pts2} 

Why is this crossover length directly connected to the configurational entropy? This can be understood following a simple thermodynamic argument.  
%We discuss the thermodynamics of the particles inside the cavity under inflence of the boundary. Suppose that the system inside the cavity was originally in a metastable state, say $\alpha$. We ask whether the metastable state $\alpha$ remains or melts recovering ergordicity.
Suppose the particles inside the cavity explore states that are very different from the reference configuration. This will allow them to sample states that have a low overlap $Q$ with the reference configuration. The free energy gained by this exploration is directly given by the Franz-Parisi free energy, $\Delta F_{-} = V(Q_{\rm glass}) v_d R^d$, where $v_d$ is the volume of the unit sphere in spatial dimension $d$. 
There is however a free energy cost to explore those states, as the radial overlap profile inside the cavity $Q(r)$ will present an interface between $Q(r = 0) \approx 0$ and $Q(r=R) \approx Q_{\rm glass}$. 
This interface in the profile of the order parameter has a free energy cost, and a simple estimate is given by $\Delta F_{+} = Y s_d R^{d-1}$, where $s_d$ is the surface area of the unit sphere in dimension $d$ and $Y$ a surface tension between two distinct glass states. 
%If the size of the cavity is very small, the particles inside the cavity are forced to stay at the same positions due to strong influence from the boundary, thus the system still remains in the state $\alpha$.  The boundary pinning effect would be quantified by an energy cost given by a surface tension term, $Y R^{d-1}$, which preventing the state $\alpha$ from melting. Physically this surface tension is intepreted as mismatch of configurations belonging to different metastable states.
In many disordered systems, the interfacial terms take a more general expression, $\Delta F_{+} = \Upsilon R^{\theta}$, where $\Upsilon$ is a generalized surface tension and the non-trivial exponent 
$\theta \leq d-1$ accounts for additional fluctuations in directions transverse to the interface.~\cite{KTW89,villain} Physically, these fluctuations arise because the system can decrease the interfacial cost by allowing the position of the interface to fluctuate and take advantage of the weakest spots.

%On the other hand, if the size of the cavity is large enough, thermal fluctuations let the state $\alpha$ melt and transform into another  state. Once the system turns into another state, it is very rare to come back to the original state $\alpha$ since there are $\mathcal{N} \approx e^{S_{\rm conf}} \approx e^{N V(Q_{\rm glass})/T}$ available states in the whole system. To go back to the initial state $\alpha$ inside the cavity, the system needs to pay $V(Q_{\rm glass}) v_d R^d$, where $v_d$ is the volume of the unit sphere in $d$-dimensions. In other words, the system ``gain'' a free energy $V(Q_{\rm glass}) v_d R^d$ by melting and recovering ergodicity. This is the entropic driving force originally discussed in Ref.~\cite{KTW89}.

The competition between exploring many states, that decreases the free energy by $\Delta F_{-}$, and the interfacial cost of an inhomogeneous overlap profile, that increases the free energy by $\Delta F_{+}$, leads to a well-defined crossover radius for the cavity where the two terms balance each other, 
\begin{equation} 
\Delta F = \Upsilon R^{\theta} - V(Q_{\rm glass}) v_d R^d=0.  
\label{eq:competition}
\end{equation}
This crossover radius defines the point-to-set correlation lengthscale $\xi_{\rm pts}$, given by
%Whether the initial state $\alpha$ survives or melts is determined by competition between the energy cost due to the surface tension, $Y R^{\theta}$, and the bulk entropic gain, $V(Q_{\rm glass}) v_d R^d$.
%When the two terms are comparable, we can define a characteristic lengthscale of a metastable state, the so-called point-to-set length, $\xi_{\rm pts}$, given by
\begin{equation}
\xi_{\rm pts} = \left( \frac{\Upsilon}{V(Q_{\rm glass}) v_d}\right)^{1/(d-\theta)}.
\label{eq:xi-V}
\end{equation}
This equation directly connects the decrease of the Franz-Parisi free energy to the growth of a spatial correlation lengthscale. It is important to notice that since $V(Q_{\rm glass})$ is unambiguously defined and can be measured in computer simulations, the same is true for the point-to-set correlation lengthscale whose existence and physical interpretation does not require any type of approximation. In particular, there is no need to assume the existence of long-lived free-energy metastable states. 

A connection between the point-to-set correlation lengthscale defined in Eq.~(\ref{eq:xi-V}) and the configurational entropy can be established by using Eq.~(\ref{eq:def_s_conf_V}) expressing the Franz-Parisi free energy $V(Q_{\rm glass}) $ as an estimate of the configurational entropy. We realize that the growth of the point-to-set correlation lengthscale as temperature decreases is equivalent to a decrease of $V(Q_{\rm glass})$, and thus to a decrease of the $S_{\rm conf}$, assuming a modest temperature dependence of $\Upsilon$.~\cite{KTW89,cammarota2009evidence} Therefore, the growth of the point-to-set correlation lengthscale is a direct real space consequence of the decrease of the configurational entropy.~\cite{KTW89,BB04} If a Kauzmann transition where $S_{\rm conf} \to 0$ \res{occurs,} then it must be accompanied by a divergence of the point-to-set correlation lengthscale, $\xi_{\rm pts} \to \infty$.  

%This equation connects the reduction of the configurational entropy with growing static correlation length or typical size of the metastable state toward the Kauzmann transition. At the Kauzmann transition, one non-erogodic metastable state extends to entire system, forming an ideal glass. 

The relation between the point-to-set lengthscale and the configurational entropy can be used both ways.~\cite{ceiling17} First, it provides a useful interpretation of the entropy crisis in terms of a diverging correlation lengthscale, as put forward in the early development of the random first order transition theory.~\cite{KTW89} We find it equally convenient to use this connection in the opposite direction and deduce from the growth of the point-to-set correlation length a quantitative determination of the variation of the configurational entropy.~\cite{ceiling17,berthier2018zero} Using the above scaling relations, the measurement of $\xi_{\rm pts}$ provides another estimate of the configurational entropy
\begin{equation}
S_{\rm conf} = N \left( \frac{\xi_0}{\xi_{\rm pts}}\right)^{d-\theta},
\label{eq:s_conf_pts}
\end{equation}
where $\xi_0$ is an unknown factor that results from conversion between entropy and lengthscale. At this stage, the value of the exponent $\theta$ is undetermined. It could be measured by comparing measurements of $S_{\rm conf}$ following Eq.~(\ref{eq:s_conf_pts}) to an independent estimate, for example from Eq.~(\ref{eq:def_s_conf_V}). The two supported values for the exponent are the simple value $\theta=d-1$,~\cite{FM07,cammarota2009numerical,cammarota2009evidence} and the renormalized value $\theta=d/2$~\cite{KTW89,lubchenko2015theory} stemming from the random interface analogy. They respectively lead to $S_{\rm conf} \sim 1/\xi_{\rm pts}$ and $S_{\rm conf} \sim 1/\xi_{\rm pts}^{d/2}$, which are equivalent in $d=2$ and not very different in $d=3$ given the relatively modest variation of the configurational entropy reported in experiments.

\subsection{Computational measurement}

\begin{figure}
\includegraphics[width=0.95\columnwidth]{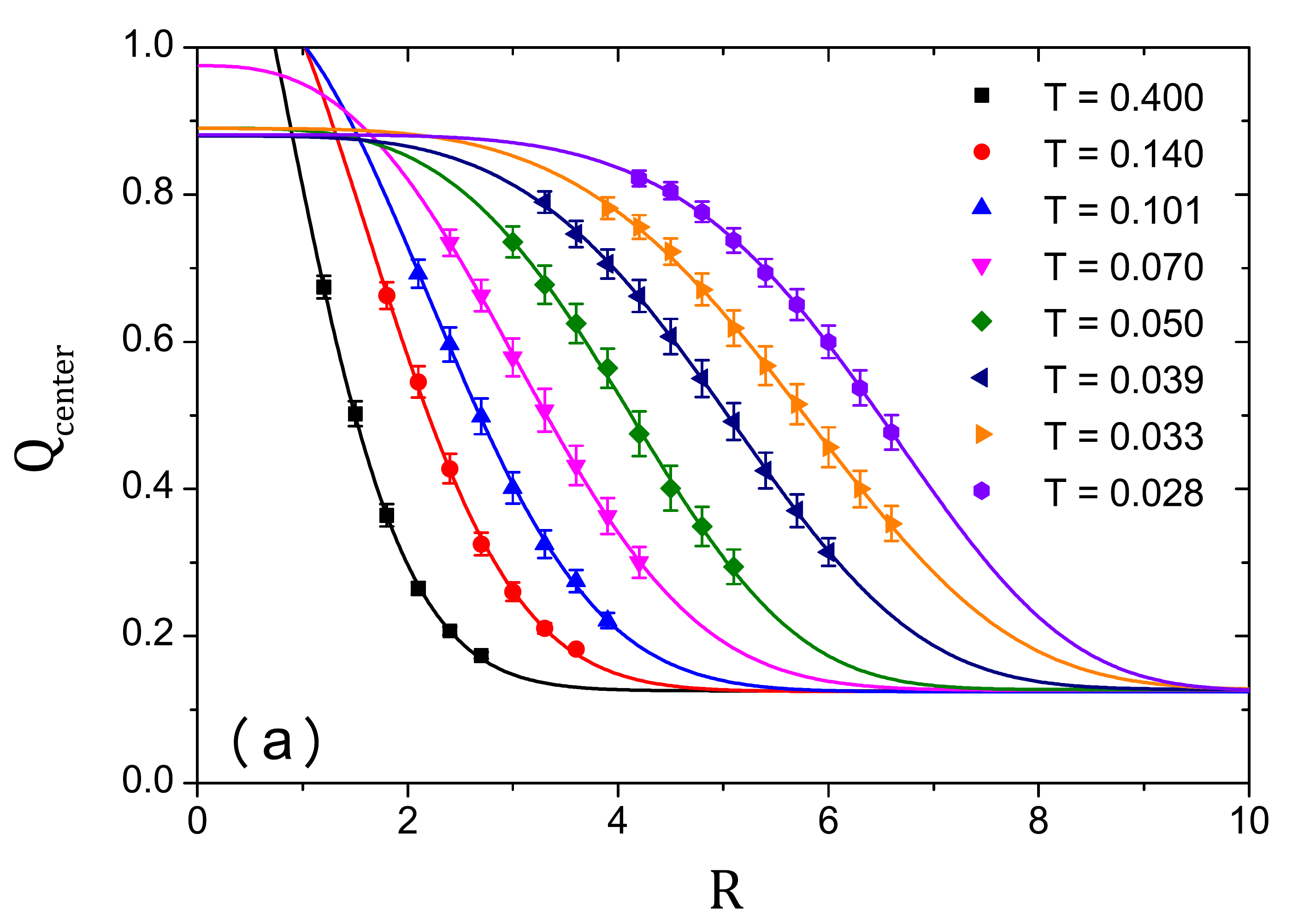}
\includegraphics[width=0.95\columnwidth]{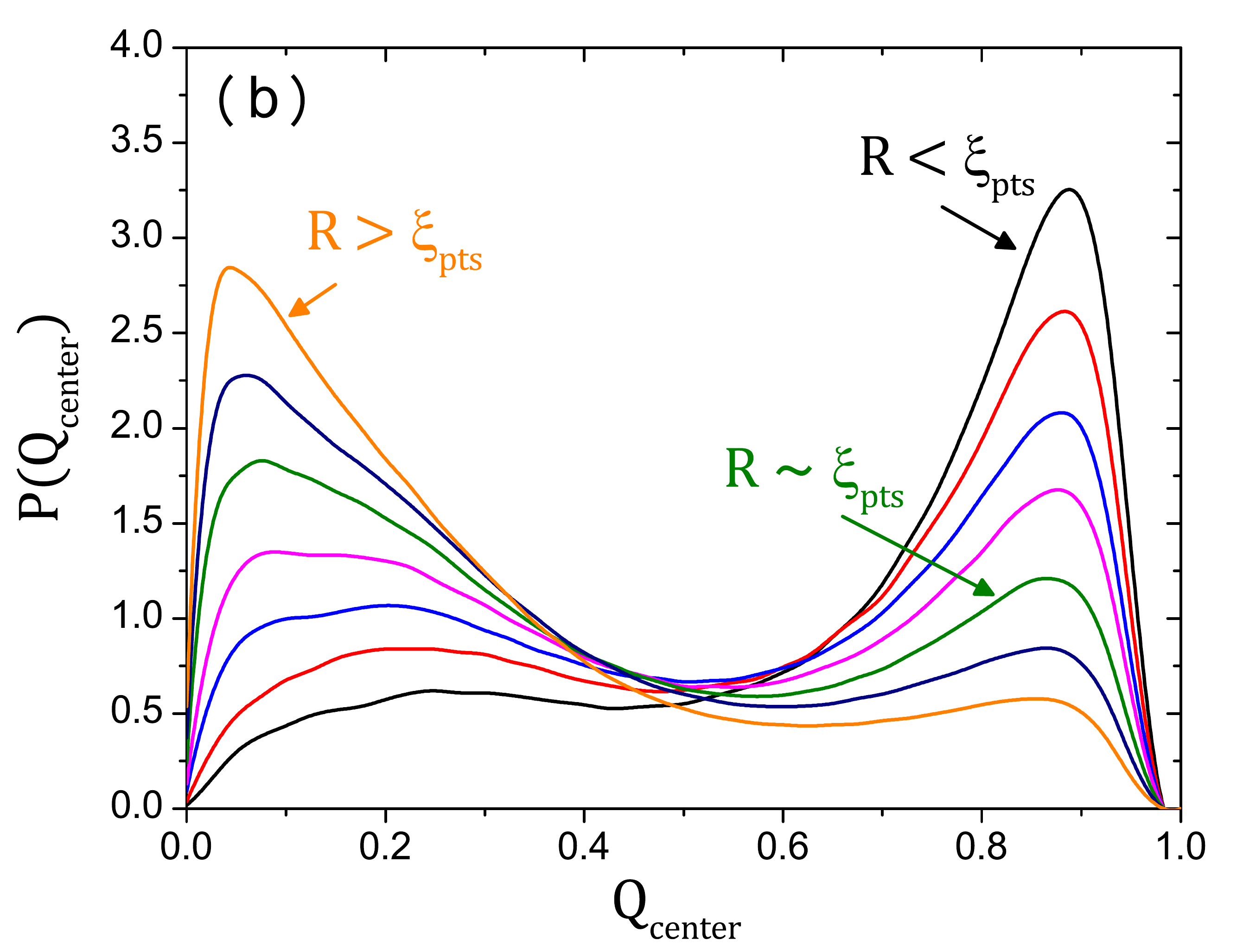}
\caption{Measurement of the point-to-set lengthscale in a $2d$ systems of polydisperse soft repulsive spheres.~\cite{berthier2018zero}
(a) Evolution of the overlap at the center of the cavity, $Q_{\rm center}$, with the cavity radius $R$ for different temperatures. 
(b) Evolution of the probability distribution of the overlap, $P(Q_{\rm center})$, with cavity radius $R$ for a given low temperature $T=0.035 \approx 0.3 T_g$.}
\label{fig:pts}
\end{figure}  

To determine the point-to-set correlation length numerically,~\cite{CGV07,BBCGV08,CGV12,hocky2012growing,berthier2016efficient} we essentially follow the theoretical construction described above and illustrated in Fig.~\ref{fig:pts_pic}. First, an equilibrium reference configuration  ${\bf r}_{\rm ref}^N$ is prepared. We define a cavity of radius $R$, centered on a randomly-chosen position in the reference configuration. We then define a configuration ${\bf r}^N$: the particles lying outside the cavity are frozen at the same positions as in the reference configuration, whereas particles inside the cavity can thermalize freely.
  
The key observable is the overlap profile $Q(r)$ between configurations ${\bf r}^N$ and ${\bf r}_{\rm ref}^N$ inside the cavity. It is numerically convenient to focus on the value of the overlap at the center of the cavity $Q_{\rm center} \equiv Q(r=0)$, which depends both on the cavity size $R$, and the temperature $T$.  
Figure~\ref{fig:pts}(a) shows the evolution of $Q_{\rm center}$ with the cavity size $R$ for polydisperse soft disks in $d=2$.~\cite{berthier2018zero} At small $R$, $Q_{\rm center} \approx Q_{\rm glass}$ meaning that the system is constrained to remain in the same state as the reference configuration. \res{The overlap is not strictly one because thermal fluctuations allow small deviations around the reference configuration ${\bf r}_{\rm ref}^N$.} At larger $R$, however, $Q_{\rm center}$ decays to a small value, which implies that the system can freely explore states that have different density profiles. The cavity size at which the transition from high to low overlap occurs determines the point-to-set lengthscale $\xi_{\rm pts}$. In practice, one can define $\xi_{\rm pts}$ when $Q_{\rm center}$ reaches a specific value, or from an empirical fitting of the whole function, such as $Q_{\rm center} \approx \exp[ -(R/\xi_{\rm pts})^b]$, where $b$ is a fitting parameter. The temperature evolution of $Q_{\rm center}(R)$ is very interesting as it directly reveals that the amorphous boundary condition constrains more strongly the interior of the cavity as the temperature decreases. Physically, it indicates that as temperature decreases, the point-to-set correlation lengthscale grows, or equivalently that the configurational entropy decreases, in virtue of Eq.~(\ref{eq:s_conf_pts}).

An interesting alternative view of the free-energy competition captured by Eq.~(\ref{eq:competition}) emerges by considering the evolution of the free-energy gain $\Delta F_{-}$ of the configuration ${\bf r}^N$ inside the cavity, as the cavity size is decreased at constant $T$. For a very large cavity, the particles in ${\bf r}^N$ are pinned at the boundaries, but those at the center of the cavity evolve as freely as in the bulk equilibrium system. Since the free-energy gain $\Delta F_{-}$ scales as $R^d$, it decreases as the cavity size decreases, making it increasingly difficult for the configuration ${\bf r}^N$ to explore other states inside cavity. As the cavity size approaches the point-to-set lengthscale, the entropic driving force to explore many states inside the cavity becomes comparable to the free energy cost $\Delta F_{+}$ of the amorphous boundary. For even smaller cavities, the system is frozen in a single state. The scenario that we have just described for the cavity is nothing but the entropy crisis predicted by the random first transition theory for the bulk system as $T \to T_K$. In other words, decreasing the cavity size for a given $T > T_K$ has an effect similar to approaching the Kauzmann transition in a bulk system. The qualitative difference is that the Kauzmann transition is a sharp thermodynamic transition happening for $N \to \infty$ in the bulk, whereas the entropy crisis in the cavity takes place for a finite system comprising $N \sim \xi_{\rm pts}^d$ particles. The crossover from small to large overlap observed around $R \sim \xi_{\rm pts}(T)$ in the profiles of Fig.~\ref{fig:pts}(a) is conceptually analogous to a Kauzmann transition rounded by the finite size of the system.~\cite{BB09}

This analogy is even more striking when the fluctuations of the overlap are recorded,~\cite{berthier2016efficient} and not only its average value. Figure~\ref{fig:pts}(b) shows the probability distribution of $Q_{\rm center}$, denoted $P(Q_{\rm center})$, for a fixed temperature as $R$ is varied. For large $R$, the distribution peaks at low values of the overlap, whereas for small $R$ it peaks near $Q_{\rm glass}$. Interestingly, at the crossover between these two regimes, $P(Q_{\rm center})$ is clearly bimodal, which is reminiscent of the distribution of the order parameter near a first-order phase transition in a finite system. These observations suggest that it is interesting to monitor the variance of these distributions, which is a measure of the susceptibility $\chi$ associated with this rounded Kauzmann transition. For a given $T$, it is found that $\chi$ has a maximum when $R = \xi_{\rm pts}$, which provides a fitting-free numerical definition of the point-to-set correlation lengthscale.~\cite{berthier2016efficient} 

Despite the conceptual simplicity of the measurements described above, it is not straightforward to obtain statistically meaningful numerical measurements of the overlap and of its fluctuations inside finite cavities. There are several reasons for this. First, to obtain a value for $\xi_{\rm pts}$ at a given temperature, one needs to analyze a range of cavity sizes that encompasses the crossover shown in Fig.~\ref{fig:pts}. For each cavity size $R$, a large number of independent cavities need to be studied, and the overlap in each individual cavity needs to be carefully monitored to ensure that its equilibrium fluctuations have been properly recorded. All in all, the number of required simulations is quite substantial. 

The second major computational obstacle naturally stems from the physics at play as $R$ is reduced. Because the confined system undergoes a finite-size analog of the Kauzmann transition, a major slowing down arises in the thermalization process. This amounts to studying an `ideal' glass transition in equilibrium conditions, an obviously daunting task. This is however possible in the present case because only a finite number of particles are contained in the cavity. This allows the use of parallel tempering (or replica exchange) methods, first developed in the context of spin glasses to overcome thermalization issues in systems with complex landscapes.~\cite{hukushima1996exchange} With these techniques, the study of a given set of parameters $(T,R)$ requires simulating a large number of copies of the system interpolating between the original system and a state point at which thermalization is fast. During the course of the simulations, exchanges between neighboring states are performed, so that each copy performs a random walk in parameter space. This method, developed in Ref.~\onlinecite{berthier2016efficient} has proven sufficiently efficient and versatile to analyze point-to-set correlations in a broad range of model systems down to very low temperetures.~\cite{ceiling17} 

\section{Perspective}

\label{sec:perspective}

We presented a short review of the configurational entropy in supercooled liquids approaching their glass transition. We first described why and how configurational entropy became a central thermodynamic quantity to describe glassy materials, both from experimental and theoretical viewpoints. We then offered our views on several paradoxes surrounding the configurational entropy. In particular, we explained that there is no reason to \res{try to avoid} an entropy crisis, that available data neither discard nor disprove its existence, and that there exists no fundamental reason, published proof, or general arguments showing that it must be avoided.  In other words, the Kauzmann transition remains a valid and useful hypothesis to interpret glass formation. We also insisted that this is still a hypothesis, but in no way a proven or necessary fact. 

The biggest paradox of all, is perhaps that the configurational entropy, which represents the key signature of the entropy crisis occurring in the modern mean-field theory of the glass transition, cannot be rigorously defined in finite dimensions as a complexity that enumerates free energy minima. We have presented several computational schemes which are meant to provide at the same time an estimate of the configurational entropy in numerical models of glass-forming liquids, and a physical interpretation that is valid in finite dimensions. 

We started with the historical method based on inherent structures, which enumerates the number of potential energy minima as well-defined, but incorrect proxies, for free energy minima. It is unclear that the inherent structure configurational entropy can in fact vanish at a Kauzmann transition, and Stillinger provided arguments that it cannot. This method is a computationally cheap method to remove a vibrational contribution to the total entropy, but it cannot be used for simple models such as hard spheres or continuously polydisperse glass-formers. 

We then showed that a generalized method elaborating on earlier ideas introduced by Frenkel and Ladd for crystals provides a better estimate of the configurational entropy, as it naturally includes both the glass mixing entropy and finite temperature anharmonicities. Additionally, the method can be applied to all types of models, including hard spheres, at a relatively cheap computational cost.   

More recent methods were developed as direct applications of the mean-field theory to computer works, which both bypass the need to mathematically define free energy minima. Free energy measurements, based on the Franz-Parisi free energy, provide an estimate for the configurational entropy that is the closest to the original mean-field definition. This method relies on the definition of a global order parameter for the glass transition, the overlap, which quantifies the similarity between pairs of configurations. Conventional methods employed in the context of equilibrium phase transitions are combined to these measurements.   

Finally, we showed that the decrease of the entropy can be given a real space interpretation in terms of a growing correlation lengthscale that is directly related to the configurational entropy. 

%\begin{figure}
%\includegraphics[width=0.95\columnwidth]{s_conf_HS.eps}
%\caption{Various estimates of the configurational entropy for three dimensional polydisperse hard spheres: Generalized Frenkel-Ladd method (GFL), Landau free energy $V(Q)$ and $\varepsilon$-coupling routes in the Franz-Parisi construction (FP), and the inverse of the point-to-set lengthscale (PTS).}
%\label{fig:all}
%\end{figure}  

%\MO{(Maybe we can mention Fig.12: All estimates are mutually consistent, despite of different conceptual settings. Because I guess reviewers of readers care about $S_{\rm conf}$ as a function of $T$ in the end, and mutual consistency among different methods.)}

This brief summary shows that there now exist conceptually solid estimates of the configurational entropy that could truly provide a direct access to the thermodynamic behavior of supercooled liquids. Given the recent progress of computer simulations to efficiently equilibrate model systems down to temperatures that are matching, and in several cases, outperforming experimental work, we feel that this is an exciting moment for glass physics, since a direct demonstration of the relevance \res{and connection to slow dynamics} of an entropy crisis and increasingly precise localizations of the putative Kauzmann transition appear possible. 

\begin{acknowledgments}
This perspective is based on two sets of lectures. The first one was given during the 2017 Boulder Summer School on ``Disordered and Frustrated systems''.
The second one was given during the 2018 Bangalore Summer School ``Entropy, information and order in soft matter''. We thank G. Biroli, P. Charbonneau, D. Coslovich, M. D. Ediger, A. Ikeda, W. Kob, K. Miyazaki, A. Ninarello, G. Parisi, G. Tarjus, and S. Yaida for collaborations and discussions on the topics discussed in this paper. We thank S. Tatsumi and O. Yamamuro for providing us with experimental data and discussions on experimental measurement of the configurational entropy. The research leading to these results has received funding from the Simons Foundation (\#454933, Ludovic Berthier). 
\end{acknowledgments}

\bibliography{biblio}

\end{document}